\documentclass[prb,twocolumn,superscriptaddress,showpacs,groupedaddress]{revtex4-2}
\usepackage[nomessages]{fp}
\usepackage{mathrsfs}
\usepackage{cancel}
\usepackage{comment}
\usepackage{enumerate}
\usepackage{amssymb}
\usepackage{amsmath}
\usepackage{braket}
\usepackage{graphicx}
\usepackage{dsfont}
\usepackage{booktabs}
\usepackage{tkz-graph}   
\usepackage{comment}
\usetikzlibrary{arrows,decorations.markings}

\newcommand{\be}{\begin{equation}}
\newcommand{\ee}{\end{equation}}
\newcommand{\ba}{\begin{aligned}}
\newcommand{\ea}{\end{aligned}}
\newcommand{\bw}{\begin{widetext}}
\newcommand{\ew}{\end{widetext}}

\newcommand{\bea}{\begin{eqnarray}}
\newcommand{\eea}{\end{eqnarray}}

\usepackage{hyperref}
\usepackage{physics}
\usepackage{dsfont}
\usepackage{amsfonts}
\usepackage{bm}
\usepackage{graphicx}
\usepackage{xcolor}
\usepackage{mathtools}
\usepackage{verbatim}
\usepackage{tikz}
\usepackage[shortlabels]{enumitem}

\usepackage{dsfont}
\usepackage{bbold}
\usepackage{tkz-graph}   
\usepackage{comment}
\usetikzlibrary{arrows,decorations.markings}

\newcommand{\graphNtwolxx}[4]
{
	\FPeval{\a}{round(#1 *(-.23),2)}
	\FPeval{\b}{round(#1 *(-.1),2)}
	\FPeval{\c}{round(#1 * (.22),2)}
	\FPeval{\d}{round(#1 *(.67),2)}
	\FPeval{\e}{round(#1 *(2)-#1*#1/2 ,2)}
	\begin{tikzpicture}[baseline={([yshift=-3pt]current bounding box.center)}]
		\GraphInit[vstyle=Classic]
		\SetVertexSimple
		\useasboundingbox (\a,\b) rectangle (\c,\d);
		\tikzset{VertexStyle/.append style={minimum size=2pt, inner sep=1pt}}
		\Vertex{a}
		\begin{scope}[decoration={markings,mark = at position 0.55 with {\arrow[thick, scale=\e]{to}}}]
			\Loop[dist=#1 cm,dir=NO,label=${\color{red}\scriptscriptstyle #2}$,style={above,thick,postaction={decorate}}](a)
		\end{scope}
		\hspace{10pt}
		\Vertex{b}
		\begin{scope}[decoration={markings,mark = at position 0.55 with {\arrow[thick, scale=\e]{to}}}]
			\Loop[dist=#1 cm,dir=NO,label=${\color{red}\scriptscriptstyle #3}$, style={above,thick,postaction={decorate}}](b)
		\end{scope}
		\hspace{10pt}
		\Vertex{c}
		\begin{scope}[decoration={markings,mark = at position 0.55 with {\arrow[thick, scale=\e]{to}}}]
			\Loop[dist=#1 cm,dir=NO,label=${\color{red}\scriptscriptstyle #4}$,style={above,thick,postaction={decorate}}](c)
		\end{scope}
	\end{tikzpicture}\hspace{20pt}
}

\newcommand{\graphN}[1]
{
	\FPeval{\a}{round(#1 *(-.23),2)}
	\FPeval{\b}{round(#1 *(-.1),2)}
	\FPeval{\c}{round(#1 * (.22),2)}
	\FPeval{\d}{round(#1 *(.67),2)}
	\FPeval{\e}{round(#1 *(2)-#1*#1/2 ,2)}
	\begin{tikzpicture}[baseline={([yshift=-3pt]current bounding box.center)}]
		\GraphInit[vstyle=Classic]
		\SetVertexSimple
		\useasboundingbox (\a,\b) rectangle (\c,\d);
		\tikzset{VertexStyle/.append style={minimum size=2pt, inner sep=1pt}}
		\Vertex{a}
		\begin{scope}[decoration={markings,mark = at position 0.55 with {\arrow[thick, scale=\e]{to}}}]
			\Loop[dist=#1 cm,dir=NO,style={thick,postaction={decorate}}](a)
		\end{scope}
	\end{tikzpicture}
}

\newcommand{\graphNparlabel}[2]
{
	\FPeval{\a}{round(#1 *(-.23),2)}
	\FPeval{\b}{round(#1 *(-.1),2)}
	\FPeval{\c}{round(#1 * (.22),2)}
	\FPeval{\d}{round(#1 *(.67),2)}
	\FPeval{\e}{round(#1 *(2)-#1*#1/2 ,2)}
	\begin{tikzpicture}[baseline={([yshift=-3pt]current bounding box.center)}]
		\GraphInit[vstyle=Classic]
		\SetVertexSimple
		\useasboundingbox (\a,\b) rectangle (\c,\d);
		\tikzset{VertexStyle/.append style={minimum size=2pt, inner sep=1pt}}
		\Vertex{a}
		\begin{scope}[decoration={markings,mark = at position 0.55 with {\arrow[thick, scale=\e]{to}}}]
			\Loop[dist=#1 cm,dir=NO,label=${\color{red}\scriptscriptstyle#2}$,style={above,thick,postaction={decorate}}](a)
		\end{scope}
	\end{tikzpicture}
}

\newcommand{\graphNred}[1]
{
	\FPeval{\a}{round(#1 *(-.23),2)}
	\FPeval{\b}{round(#1 *(-.1),2)}
	\FPeval{\c}{round(#1 * (.22),2)}
	\FPeval{\d}{round(#1 *(.67),2)}
	\FPeval{\e}{round(#1 *(2)-#1*#1/2 ,2)}
	\begin{tikzpicture}[baseline={([yshift=-3pt]current bounding box.center)}]
		\GraphInit[vstyle=Classic]
		\SetVertexSimple
		\useasboundingbox (\a,\b) rectangle (\c,\d);
		\tikzset{VertexStyle/.append style={minimum size=2pt, inner sep=1pt,color=red}}
		\Vertex{a}
		\begin{scope}[decoration={markings,mark = at position 0.55 with {\arrow[thick, scale=\e]{to}}}]
			\Loop[dist=#1 cm,dir=NO,style={thick,color=red,postaction={decorate}}](a)
		\end{scope}
	\end{tikzpicture}
}

\newcommand{\graphNtwofin}[3]
{
	\FPeval{\a}{round(#1 *(-.23),2)}
	\FPeval{\b}{round(#1 *(-.1),2)}
	\FPeval{\c}{round(#1 * (.22),2)}
	\FPeval{\d}{round(#1 *(.67),2)}
	\FPeval{\e}{round(#1 *(2)-#1*#1/2 ,2)}
	\hspace{2pt}\begin{tikzpicture}[baseline={([yshift=-3pt]current bounding box.center)}]
		\GraphInit[vstyle=Classic]
		\SetVertexSimple
		\useasboundingbox (\a,\b) rectangle (\c,\d);
		\tikzset{VertexStyle/.append style={minimum size=2pt, inner sep=1pt}}
		\Vertex{a}
		\begin{scope}[decoration={markings,mark = at position 0.55 with {\arrow[thick, scale=\e]{to}}}]
			\Loop[dist=#1 cm,dir=NO,label=$\scriptscriptstyle {\color{red}#2}$,style={above,thick,postaction={decorate}}](a)
		\end{scope}
		\hspace{10pt}
		\Vertex{b}
		\begin{scope}[decoration={markings,mark = at position 0.55 with {\arrow[thick, scale=\e]{to}}}]
			\Loop[dist=#1 cm,dir=NO,label=$\scriptscriptstyle{\color{red}#3}$,style={above,thick,postaction={decorate}}](b)
		\end{scope}
	\end{tikzpicture}\hspace{12pt}
}

\newcommand{\graphNdeuxred}[1]
{
	\FPeval{\a}{round(#1 *(-.65),2)}
	\FPeval{\b}{round(#1 *(-.265),3)}
	\FPeval{\c}{round(#1 * (.66),2)}
	\FPeval{\d}{round(#1 *(.24),2)}
	\FPeval{\e}{round(#1 *(2)-#1*#1/2 ,2)}
	\begin{tikzpicture}[baseline={([yshift=-3pt]current bounding box.center)}]
		\GraphInit[vstyle=Classic]
		\SetVertexSimple
		\tikzset{VertexStyle/.append style={minimum size=2pt, inner sep=1pt, color=red}}
		\Vertex{a}
		\begin{scope}[decoration={markings,mark = at position 0.55 with {\arrow[thick, scale=\e]{to}}}]
			\Loop[dist=#1 cm,dir=WE,style={thick,color=red,postaction={decorate}}](a)
			\Loop[dist=#1 cm,dir=EA,style={thick,color=red,postaction={decorate}}](a)
		\end{scope}
	\end{tikzpicture}
}

\newcommand{\graphNdeux}[1]
{
	\FPeval{\a}{round(#1 *(-.65),2)}
	\FPeval{\b}{round(#1 *(-.265),3)}
	\FPeval{\c}{round(#1 * (.66),2)}
	\FPeval{\d}{round(#1 *(.24),2)}
	\FPeval{\e}{round(#1 *(2)-#1*#1/2 ,2)}
	\begin{tikzpicture}[baseline={([yshift=-3pt]current bounding box.center)}]
		\GraphInit[vstyle=Classic]
		\SetVertexSimple
		\tikzset{VertexStyle/.append style={minimum size=2pt, inner sep=1pt}}
		\Vertex{a}
		\begin{scope}[decoration={markings,mark = at position 0.55 with {\arrow[thick, scale=\e]{to}}}]
			\Loop[dist=#1 cm,dir=WE,style={thick,postaction={decorate}}](a)
			\Loop[dist=#1 cm,dir=EA,style={thick,postaction={decorate}}](a)
		\end{scope}
	\end{tikzpicture}
}

\newcommand{\graphNtroislabel}[4]
{
	\FPeval{\a}{round(#1 *(-.5),2)}
	\FPeval{\b}{round(#1 *(-.53),2)}
	\FPeval{\c}{round(#1 * (.62),2)}
	\FPeval{\d}{round(#1 *(.68),2)}
	\FPeval{\e}{round(#1 *(2)-#1*#1/2 ,2)}\hspace{6pt}
	\begin{tikzpicture}[baseline={([yshift=-3pt]current bounding box.center)}]
		\GraphInit[vstyle=Classic]
		\SetVertexSimple
		\useasboundingbox (\a,\b) rectangle (\c,\d);
		\tikzset{VertexStyle/.append style={minimum size=2pt, inner sep=1pt}}
		\Vertex{a}
		\begin{scope}[decoration={markings,mark = at position 0.55 with {\arrow[thick, scale=\e]{to}}}]
			\Loop[dist=#1 cm,label=${\color{red}\scriptscriptstyle#2}$,dir=NO,style={above,thick,postaction={decorate}}](a)
			\Loop[dist=#1 cm,label=${\color{red}\scriptscriptstyle#3}$,dir=EA,style={right,thick,postaction={decorate}}](a)
		\Loop[dist=#1 cm,label=${\color{red}\scriptscriptstyle#4}$,dir=SOWE,style={left,thick,postaction={decorate}}](a)\end{scope}
	\end{tikzpicture}\hspace{6pt}
}

\newcommand{\graphNquatrlabel}[5]
{
	\FPeval{\a}{round(#1 *(-.5),2)}
	\FPeval{\b}{round(#1 *(-.53),2)}
	\FPeval{\c}{round(#1 * (.62),2)}
	\FPeval{\d}{round(#1 *(.68),2)}
	\FPeval{\e}{round(#1 *(2)-#1*#1/2 ,2)}\hspace{6pt}
	\begin{tikzpicture}[baseline={([yshift=-3pt]current bounding box.center)}]
		\GraphInit[vstyle=Classic]
		\SetVertexSimple
		\useasboundingbox (\a,\b) rectangle (\c,\d);
		\tikzset{VertexStyle/.append style={minimum size=2pt, inner sep=1pt}}
		\Vertex{a}
		\begin{scope}[decoration={markings,mark = at position 0.55 with {\arrow[thick, scale=\e]{to}}}]
			\Loop[dist=#1 cm,label=${\color{red}\scriptscriptstyle#2}$,dir=NO,style={above,thick,postaction={decorate}}](a)
			\Loop[dist=#1 cm,label=${\color{red}\scriptscriptstyle#3}$,dir=EA,style={right,thick,postaction={decorate}}](a)
		\Loop[dist=#1 cm,label=${\color{red}\scriptscriptstyle#4}$,dir=SO,style={left,thick,postaction={decorate}}](a)
		\Loop[dist=#1 cm,label=${\color{red}\scriptscriptstyle#5}$,dir=WE,style={left,thick,postaction={decorate}}](a)	
\end{scope}
	\end{tikzpicture}
	\hspace{6pt}
}

\newcommand{\graphFdeuxred}[1]
{
	\FPeval{\a}{round(#1 *(-.04),3)}
	\FPeval{\b}{round(#1 *(-.45),2)}
	\FPeval{\c}{round(#1 * (1.05),3)}
	\FPeval{\d}{round(#1 *(.45),2)}
	\FPeval{\e}{round(#1 *(2)-#1*#1/2 ,2)}  
	\begin{tikzpicture}[baseline={([yshift=-3pt]current bounding box.center)}]
		\GraphInit[vstyle=Classic]
		\SetGraphUnit{#1} 
		\SetVertexSimple
		\useasboundingbox (\a,\b) rectangle (\c,\d);
		\tikzset{VertexStyle/.append style={minimum size=2pt, color=red,inner sep=1pt}}
		\Vertex{a}
		\EA(a){b}
		\begin{scope}[decoration={markings,mark = at position 0.55 with {\arrow[thick, scale=\e]{to}}}]
			\tikzset{EdgeStyle/.style = {postaction={decorate},color=red,bend left=75}}
			\Edge(a)(b)
		\end{scope}
		\begin{scope}[decoration={markings,mark = at position 0.55 with {\arrow[thick, scale=\e]{to}}}]
			\tikzset{EdgeStyle/.style = {postaction={decorate},color=red,bend left=75}}
			\Edge(b)(a)
		\end{scope}
	\end{tikzpicture}
}

\newcommand{\graphFdeuxfin}[3]
{
	\FPeval{\a}{round(#1 *(-.04),3)}
	\FPeval{\b}{round(#1 *(-.45),2)}
	\FPeval{\c}{round(#1 * (1.05),3)}
	\FPeval{\d}{round(#1 *(.45),2)}
	\FPeval{\e}{round(#1 *(2)-#1*#1/2 ,2)}  \hspace{2pt}
	\begin{tikzpicture}[baseline={([yshift=-3pt]current bounding box.center)}]
		\GraphInit[vstyle=Classic]
		\SetGraphUnit{#1} 
		\SetVertexSimple
		\useasboundingbox (\a,\b) rectangle (\c,\d);
		\tikzset{VertexStyle/.append style={minimum size=2pt, inner sep=1pt}}
		\Vertex{a}
		\EA(a){b}
		\begin{scope}[decoration={markings,mark = at position 0.55 with {\arrow[thick, scale=\e]{to}}}]
			\tikzset{EdgeStyle/.style = {postaction={decorate},bend right=75}}
			\Edge[label=$\scriptscriptstyle{\color{red}#2}$,style={sloped,below, draw opacity=1 ,fill opacity=0, text opacity=1}](a)(b)
		\end{scope}
		\begin{scope}[decoration={markings,mark = at position 0.55 with {\arrow[thick, scale=\e]{to}}}]
			\tikzset{EdgeStyle/.style = {postaction={decorate},bend left=75}}
			\Edge[label=$\scriptscriptstyle{\color{red}#3}$,style={sloped,above, draw opacity=1 ,fill opacity=0, text opacity=1}](a)(b)
		\end{scope}
	\end{tikzpicture}\hspace{2pt}
}

\newcommand{\graphFdeuxr}[1]
{
	\FPeval{\a}{round(#1 *(-.04),3)}
	\FPeval{\b}{round(#1 *(-.45),2)}
	\FPeval{\c}{round(#1 * (1.05),3)}
	\FPeval{\d}{round(#1 *(.45),2)}
	\FPeval{\e}{round(#1 *(2)-#1*#1/2 ,2)}  
	\begin{tikzpicture}[baseline={([yshift=-3pt]current bounding box.center)}]
		\GraphInit[vstyle=Classic]
		\SetGraphUnit{#1} 
		\SetVertexSimple
		\useasboundingbox (\a,\b) rectangle (\c,\d);
		\tikzset{VertexStyle/.append style={minimum size=2pt, inner sep=1pt}}
		\Vertex{a}
		\EA(a){b}
		\begin{scope}[decoration={markings,mark = at position 0.55 with {\arrow[thick, scale=\e]{to}}}]
			\tikzset{EdgeStyle/.style = {postaction={decorate},bend right=75}}
			\Edge(a)(b)
		\end{scope}
		\begin{scope}[decoration={markings,mark = at position 0.55 with {\arrow[thick, scale=\e]{to}}}]
			\tikzset{EdgeStyle/.style = {postaction={decorate},bend left=75}}
			\Edge(a)(b)
		\end{scope}
	\end{tikzpicture}
}

\newcommand{\graphFdeux}[1]
{
	\FPeval{\a}{round(#1 *(-.04),3)}
	\FPeval{\b}{round(#1 *(-.45),2)}
	\FPeval{\c}{round(#1 * (1.05),3)}
	\FPeval{\d}{round(#1 *(.45),2)}
	\FPeval{\e}{round(#1 *(2)-#1*#1/2 ,2)}  
	\begin{tikzpicture}[baseline={([yshift=-3pt]current bounding box.center)}]
		\GraphInit[vstyle=Classic]
		\SetGraphUnit{#1} 
		\SetVertexSimple
		\useasboundingbox (\a,\b) rectangle (\c,\d);
		\tikzset{VertexStyle/.append style={minimum size=2pt, inner sep=1pt}}
		\Vertex{a}
		\EA(a){b}
		\begin{scope}[decoration={markings,mark = at position 0.55 with {\arrow[thick, scale=\e]{to}}}]
			\tikzset{EdgeStyle/.style = {postaction={decorate},bend left=75}}
			\Edge(b)(a)
			\Edge(a)(b)
		\end{scope}
	\end{tikzpicture}
}

\newcommand{\graphNpointFdeuxllabel}[4]
{
	\FPeval{\a}{round(#1 *(-.04),3)}
	\FPeval{\b}{round(#1 *(-2/5),2)}
	\FPeval{\d}{round(#1 *(.4),2)}
	\FPeval{\e}{round(#1 *(2)-#1*#1/2 ,2)}
	\FPeval{\c}{round(#1 * (1.59)+\e*(.07),2)}\hspace{14pt}
	\begin{tikzpicture}[baseline={([yshift=-3pt]current bounding box.center)}]
		\GraphInit[vstyle=Classic]
		\SetGraphUnit{#1}   
		\SetVertexSimple
		\useasboundingbox (\a,\b) rectangle (\c,\d);  
		\tikzset{VertexStyle/.append style={minimum size=2pt, inner sep=1pt}}
		\Vertex{a}
		\EA(a){b}
		\begin{scope}[decoration={markings,mark = at position 0.55 with {\arrow[thick, scale=\e]{to}}}]
			\tikzset{EdgeStyle/.style = {postaction={decorate},bend left=75}}
			\Edge[label=$\color{red}\scriptscriptstyle#2$,style={sloped,below, draw opacity=1 ,fill opacity=0, text opacity=1}](b)(a)
			\Edge[label=$\color{red}\scriptscriptstyle#3$,style={sloped,above, draw opacity=1 ,fill opacity=0, text opacity=1}](a)(b)
			\tikzset{EdgeStyle/.style = {postaction={decorate}}}
			\Loop[dist=#1 cm,dir=WE,label=$\color{red}\scriptscriptstyle#4$,style={left,thick,postaction={decorate}}](a)
		\end{scope}
	\end{tikzpicture}\hspace{-5pt}
}

\newcommand{\graphNnodefin}[4]
{
	\FPeval{\a}{round(#1 *(-.04),3)}
	\FPeval{\b}{round(#1 *(-2/5),2)}
	\FPeval{\d}{round(#1 *(.4),2)}
	\FPeval{\e}{round(#1 *(2)-#1*#1/2 ,2)}
	\FPeval{\c}{round(#1 * (1.59)+\e*(.07),2)}
	\begin{tikzpicture}[baseline={([yshift=-3pt]current bounding box.center)}]
		\hspace{16pt}
		\GraphInit[vstyle=Classic]
		\SetGraphUnit{#1}   
		\SetVertexSimple
		\useasboundingbox (\a,\b) rectangle (\c,\d);  
		\tikzset{VertexStyle/.append style={color=#2,minimum size=2pt, inner sep=1pt}}
		\Vertex{a}
		\begin{scope}[decoration={markings,mark = at position 0.55 with {\arrow[thick, scale=\e]{to}}}]
			\tikzset{EdgeStyle/.style = {postaction={decorate}}}
			\Loop[dist=#1 cm,color=#2,dir=WE,label=$\scriptscriptstyle{\color{red}#3}$,style={left,thick,postaction={decorate}}](a)
			\Loop[dist=#1 cm,color=#2,dir=EA,label=$\scriptscriptstyle{\color{red}#4}$,style={right,thick,postaction={decorate}}](a)
		\end{scope}
	\end{tikzpicture}\hspace{8pt}
}

\newcommand{\graphNnodelabel}[3]
{
	\FPeval{\a}{round(#1 *(-.04),3)}
	\FPeval{\b}{round(#1 *(-2/5),2)}
	\FPeval{\d}{round(#1 *(.4),2)}
	\FPeval{\e}{round(#1 *(2)-#1*#1/2 ,2)}
	\FPeval{\c}{round(#1 * (1.59)+\e*(.07),2)}\hspace{13pt}
	\begin{tikzpicture}[baseline={([yshift=-3pt]current bounding box.center)}]
		\GraphInit[vstyle=Classic]
		\SetGraphUnit{#1}   
		\SetVertexSimple
		\useasboundingbox (\a,\b) rectangle (\c,\d);  
		\tikzset{VertexStyle/.append style={minimum size=2pt, inner sep=1pt}}
		\Vertex{a}
		\begin{scope}[decoration={markings,mark = at position 0.55 with {\arrow[thick, scale=\e]{to}}}]
			\tikzset{EdgeStyle/.style = {postaction={decorate}}}
			\Loop[dist=#1 cm,dir=WE,label=${\color{red}\scriptscriptstyle#2}$,style={left,thick,postaction={decorate}}](a)
			\Loop[dist=#1 cm,dir=EA,label=${\color{red}\scriptscriptstyle#3}$,style={right,thick,postaction={decorate}}](a)
		\end{scope}
	\end{tikzpicture}
}

\newcommand{\graphNpointFdeuxlrrlabel}[4]
{
	\FPeval{\a}{round(#1 *(-.04),3)}
	\FPeval{\b}{round(#1 *(-2/5),2)}
	\FPeval{\d}{round(#1 *(.4),2)}
	\FPeval{\e}{round(#1 *(2)-#1*#1/2 ,2)}
	\FPeval{\c}{round(#1 * (1.59)+\e*(.07),2)}\hspace{14pt}
	\begin{tikzpicture}[baseline={([yshift=-3pt]current bounding box.center)}]
		\GraphInit[vstyle=Classic]
		\SetGraphUnit{#1}   
		\SetVertexSimple
		\useasboundingbox (\a,\b) rectangle (\c,\d);  
		\tikzset{VertexStyle/.append style={minimum size=2pt, inner sep=1pt}}
		\Vertex{a}
		\EA(a){b}
		\begin{scope}[decoration={markings,mark = at position 0.55 with {\arrow[thick, scale=\e]{to}}}]
			\tikzset{EdgeStyle/.style = {postaction={decorate},bend left=75}}
			\Edge[label=${\color{red}\scriptscriptstyle#2}$,style={sloped,below, draw opacity=1 ,fill opacity=0, text opacity=1}](b)(a)
			\tikzset{EdgeStyle/.style = {postaction={decorate},bend right=75}}
			\Edge[label=${\color{red}\scriptscriptstyle#3}$,style={sloped,above, draw opacity=1 ,fill opacity=0, text opacity=1}](b)(a)
			\tikzset{EdgeStyle/.style = {postaction={decorate}}}
			\Loop[dist=#1 cm,dir=WE,label=${\color{red}\scriptscriptstyle#4}$,style={left,thick,postaction={decorate}}](a)
		\end{scope}
	\end{tikzpicture}\hspace{-5pt}
}

\newcommand{\graphNpointFdeuxlrlabel}[4]
{
	\FPeval{\a}{round(#1 *(-.04),3)}
	\FPeval{\b}{round(#1 *(-2/5),2)}
	\FPeval{\d}{round(#1 *(.4),2)}
	\FPeval{\e}{round(#1 *(2)-#1*#1/2 ,2)}
	\FPeval{\c}{round(#1 * (1.59)+\e*(.07),2)}\hspace{3pt}
	\begin{tikzpicture}[baseline={([yshift=-3pt]current bounding box.center)}]
		\GraphInit[vstyle=Classic]
		\SetGraphUnit{#1}   
		\SetVertexSimple
		\useasboundingbox (\a,\b) rectangle (\c,\d);  
		\tikzset{VertexStyle/.append style={minimum size=2pt, inner sep=1pt}}
		\Vertex{a}
		\EA(a){b}
		\begin{scope}[decoration={markings,mark = at position 0.55 with {\arrow[thick, scale=\e]{to}}}]
			\tikzset{EdgeStyle/.style = {postaction={decorate},bend left=75}}
			\Edge[label=${\color{red}\scriptscriptstyle#2}$,style={sloped,below, draw opacity=1 ,fill opacity=0, text opacity=1}](b)(a)
			\tikzset{EdgeStyle/.style = {postaction={decorate},bend right=75}}
			\Edge[label=${\color{red}\scriptscriptstyle#3}$,style={sloped,above, draw opacity=1 ,fill opacity=0, text opacity=1}](b)(a)
			\tikzset{EdgeStyle/.style = {postaction={decorate}}}
			\Loop[dist=#1 cm,dir=EA,label=${\color{red}\scriptscriptstyle#4}$,style={right,thick,postaction={decorate}}](b)
		\end{scope}
	\end{tikzpicture}\hspace{2pt}
}

\newcommand{\graphlineintro}[3]
{
	\FPeval{\a}{round(#1 *(-1),2)}
	\FPeval{\b}{round(#1 *(-1.05),2)}
	\FPeval{\c}{round(#1 * (1.1),2)}
	\FPeval{\d}{round(#1 *(1.05),2)}
	\FPeval{\e}{round(#1 *(5)-2*#1*#1 ,2)} 
	\begin{tikzpicture}[baseline={([yshift=-3pt]current bounding box.center)}]
		\GraphInit[vstyle=Classic]
		\SetGraphUnit{#1}   
		\SetVertexSimple
		\useasboundingbox (\a,\b) rectangle (\c,\d); 
		\tikzset{VertexStyle/.append style={color=#3,minimum size=2pt, inner sep=1pt}}
		\begin{scope}[decoration={markings,mark = at position 0.5 with {\arrow[thick, scale=\e]{to}}}]
			\tikzset{EdgeStyle/.style = {postaction={decorate},bend left}}
			\Vertices{circle}{a,b} 
			\Edge[label=$\scriptstyle{\color{red} #2}$,style={color=#3,sloped,above, draw opacity=1 ,fill opacity=0, text opacity=1}](b)(a)
		\end{scope}
	\end{tikzpicture}
}

\newcommand{\graphlinefin}[2]
{
	\FPeval{\a}{round(#1 *(-1),2)}
	\FPeval{\b}{round(#1 *(-1.05),2)}
	\FPeval{\c}{round(#1 * (1.1),2)}
	\FPeval{\d}{round(#1 *(1.05),2)}
	\FPeval{\e}{round(#1 *(5)-2*#1*#1 ,2)} 
	\begin{tikzpicture}[baseline={([yshift=-3pt]current bounding box.center)}]
		\GraphInit[vstyle=Classic]
		\SetGraphUnit{#1}   
		\SetVertexSimple
		\useasboundingbox (\a,\b) rectangle (\c,\d); 
		\tikzset{VertexStyle/.append style={minimum size=2pt, inner sep=1pt}}
		\begin{scope}[decoration={markings,mark = at position 0.5 with {\arrow[thick, scale=\e]{to}}}]
			\tikzset{EdgeStyle/.style = {postaction={decorate},bend left}}
			\Vertices{circle}{a,b} 
			\Edge[label=$\scriptstyle{\color{red} #2}$,style={sloped,above, draw opacity=1 ,fill opacity=0, text opacity=1}](b)(a)
		\end{scope}
	\end{tikzpicture}
}

\newcommand{\graphline}[1]
{
	\FPeval{\a}{round(#1 *(-1),2)}
	\FPeval{\b}{round(#1 *(-1.05),2)}
	\FPeval{\c}{round(#1 * (1.1),2)}
	\FPeval{\d}{round(#1 *(1.05),2)}
	\FPeval{\e}{round(#1 *(5)-2*#1*#1 ,2)} 
	\begin{tikzpicture}[baseline={([yshift=-3pt]current bounding box.center)}]
		\GraphInit[vstyle=Classic]
		\SetGraphUnit{#1}   
		\SetVertexSimple
		\useasboundingbox (\a,\b) rectangle (\c,\d); 
		\tikzset{VertexStyle/.append style={minimum size=2pt, inner sep=1pt}}
		\begin{scope}[decoration={markings,mark = at position 0.5 with {\arrow[thick, scale=\e]{to}}}]
			\tikzset{EdgeStyle/.style = {postaction={decorate},bend left}}
			\Vertices{circle}{a,b} 
			\Edge[label=$\scriptstyle{\color{red} a}$,style={sloped,above, draw opacity=1 ,fill opacity=0, text opacity=1}](b)(a)
		\end{scope}
	\end{tikzpicture}
}

\newcommand{\graphFktroislabel}[4]
{
	\FPeval{\a}{round(#1 *(-1),2)}
	\FPeval{\b}{round(#1 *(-1.05),2)}
	\FPeval{\c}{round(#1 * (1.1),2)}
	\FPeval{\d}{round(#1 *(1.05),2)}
	\FPeval{\e}{round(#1 *(5)-2*#1*#1 ,2)} \hspace{6pt}
	\begin{tikzpicture}[baseline={([yshift=-3pt]current bounding box.center)}]
		\GraphInit[vstyle=Classic]
		\SetGraphUnit{#1}   
		\SetVertexSimple
		\useasboundingbox (\a,\b) rectangle (\c,\d); 
		\tikzset{VertexStyle/.append style={minimum size=2pt, inner sep=1pt}}
		\begin{scope}[decoration={markings,mark = at position 0.55 with {\arrow[thick, scale=\e]{to}}}]
			\tikzset{EdgeStyle/.style = {postaction={decorate},bend right}}
			\Vertices{circle}{a,b,c} 
			\Edge[label=${\color{red}\scriptscriptstyle#2}$,style={sloped,above, draw opacity=1 ,fill opacity=0, text opacity=1}](a)(b)
			\Edge[label=${\color{red}\scriptscriptstyle#3}$,style={sloped,below, draw opacity=1 ,fill opacity=0, text opacity=1}](b)(c)
			\tikzset{EdgeStyle/.style = {postaction={decorate},bend left}}
			\Edge[label=${\color{red}\scriptscriptstyle#4}$,style={sloped,below, draw opacity=1 ,fill opacity=0, text opacity=1}](a)(c)
		\end{scope}
	\end{tikzpicture}
}

\newcommand{\graphFtroisllabel}[4]
{
	\FPeval{\a}{round(#1 *(-1),2)}
	\FPeval{\b}{round(#1 *(-1.05),2)}
	\FPeval{\c}{round(#1 * (1.1),2)}
	\FPeval{\d}{round(#1 *(1.05),2)}
	\FPeval{\e}{round(#1 *(5)-2*#1*#1 ,2)} \hspace{5pt}
	\begin{tikzpicture}[baseline={([yshift=-3pt]current bounding box.center)}]
		\GraphInit[vstyle=Classic]
		\SetGraphUnit{#1}   
		\SetVertexSimple
		\useasboundingbox (\a,\b) rectangle (\c,\d); 
		\tikzset{VertexStyle/.append style={minimum size=2pt, inner sep=1pt}}
		\begin{scope}[decoration={markings,mark = at position 0.55 with {\arrow[thick, scale=\e]{to}}}]
			\tikzset{EdgeStyle/.style = {postaction={decorate},bend right}}
			\Vertices{circle}{a,b,c} 
			\Edge[label=${\color{red}\scriptscriptstyle#2}$,style={sloped,above, draw opacity=1 ,fill opacity=0, text opacity=1}](a)(b)
			\Edge[label=${\color{red}\scriptscriptstyle#3}$, style={sloped,below, draw opacity=1 ,fill opacity=0, text opacity=1}](b)(c)
			\tikzset{EdgeStyle/.style = {postaction={decorate},bend right}}
			\Edge[label=${\color{red}\scriptscriptstyle#4}$, style={sloped,below, draw opacity=1 ,fill opacity=0, text opacity=1}](c)(a)
		\end{scope}
	\end{tikzpicture}
}

\newcommand{\graphNFdeuxllabel}[4]
{
	\FPeval{\e}{round(#1 *(1.2) ,2)}
	\FPeval{\f}{round(#1 *(4)-#1*#1 ,2)}  
	\graphNparlabel{#1}{#2}
	\hspace{\f pt}
	\begin{tikzpicture}[baseline={([yshift=-3pt]current bounding box.center)}]
		\GraphInit[vstyle=Classic]
		\SetGraphUnit{#1} 
		\SetVertexSimple
		\tikzset{VertexStyle/.append style={minimum size=2pt, inner sep=1pt}}
		\Vertex{a}
		\EA(a){b}
		\begin{scope}[decoration={markings,mark = at position 0.55 with {\arrow[thick, scale=\e]{to}}}]
			\tikzset{EdgeStyle/.style = {postaction={decorate},bend left=75}}
			\Edge[label=${\color{red}\scriptscriptstyle#3}$,style={sloped,below, draw opacity=1 ,fill opacity=0, text opacity=1}](b)(a)
			\Edge[label=${\color{red}\scriptscriptstyle#4}$,style={sloped,above, draw opacity=1 ,fill opacity=0, text opacity=1}](a)(b)
		\end{scope}
	\end{tikzpicture}
}

\newcommand{\graphNFdeuxfin}[3]
{
	\FPeval{\e}{round(#1 *(1.8) ,3)}
	\FPeval{\f}{round(#1 *(4)-#1*#1 ,2)}  
	\hspace{\f pt}\hspace{0pt}
	\begin{tikzpicture}[baseline={([yshift=-3pt]current bounding box.center)}]
		\GraphInit[vstyle=Classic]
		\SetGraphUnit{#1} 
		\SetVertexSimple
		\tikzset{VertexStyle/.append style={minimum size=2pt, inner sep=1pt}}
		\Vertex{a}
		\EA(a){b}
		\begin{scope}[decoration={markings,mark = at position 0.55 with {\arrow[thick, scale=\e]{to}}}]
			\tikzset{EdgeStyle/.style = {postaction={decorate},bend right=75}}
			\Edge[label=$\scriptscriptstyle{\color{red}#2}$,style={sloped,below, draw opacity=1 ,fill opacity=0, text opacity=1}](a)(b)
			\tikzset{EdgeStyle/.style = {postaction={decorate},bend right=75}}
			\Edge[label=$\scriptscriptstyle{\color{red}#3}$,style={sloped,above, draw opacity=1 ,fill opacity=0, text opacity=1}](b)(a)
		\end{scope}
	\end{tikzpicture}\hspace{2pt}
}

\newcommand{\graphNFdeuxlrlabel}[4]
{
	\FPeval{\e}{round(#1 *(1.2) ,2)}
	\FPeval{\f}{round(#1 *(4)-#1*#1 ,2)}  
	\graphNparlabel{#1}{#2}
	\hspace{\f pt}
	\begin{tikzpicture}[baseline={([yshift=-3pt]current bounding box.center)}]
		\GraphInit[vstyle=Classic]
		\SetGraphUnit{#1} 
		\SetVertexSimple
		\tikzset{VertexStyle/.append style={minimum size=2pt, inner sep=1pt}}
		\Vertex{a}
		\EA(a){b}
		\begin{scope}[decoration={markings,mark = at position 0.55 with {\arrow[thick, scale=\e]{to}}}]
			\tikzset{EdgeStyle/.style = {postaction={decorate},bend left=75}}
			\Edge[label=${\color{red}\scriptscriptstyle #3}$,style={sloped,below, draw opacity=1 ,fill opacity=0, text opacity=1}](b)(a)
			\tikzset{EdgeStyle/.style = {postaction={decorate},bend right=75}}
			\Edge[label=${\color{red}\scriptscriptstyle #4}$, style={sloped,above, draw opacity=1 ,fill opacity=0, text opacity=1}](b)(a)
		\end{scope}
	\end{tikzpicture}
}

\begin{document}
	
\title{$\nu$-QSSEP: A toy model for entanglement spreading in stochastic diffusive quantum systems
}
\date{\today}
\author{Vincenzo Alba}
\affiliation{Dipartimento di Fisica dell'Universit\`a di Pisa and INFN, Sezione di Pisa, I-56127 Pisa, Italy}
	
\begin{abstract}
	We investigate out-of-equilibrium entanglement dynamics in  a generalization of 
	the so-called $QSSEP$ model, which is a free-fermion chain with stochastic in space and time hopping amplitudes. 
	In our setup, the noisy amplitudes are spatially-modulated satisfying a $\nu$-site translation 
	invariance but retaining their randomness in time. 
	For each noise realization, the dynamics preserves Gaussianity, which allows to obtain noise-averaged 
	entanglement-related quantities. 
	The statistics of the steady-state correlators satisfy nontrivial relationships 
	that are of topological nature. They reflect the Haar invariance under multiplication with 
	structured momentum-dependent random $SU(\nu)$ matrices. We discuss in detail the case with 
	$\nu=1$ and $\nu=2$.  For $\nu=1$, i.e., spatially homogeneous noise we show that the entanglement 
	dynamics is describable by a stochastic generalization of the quasiparticle picture. 
	Precisely, entanglement is propagated by pairs of quasiparticles. The entanglement 
	content of the pairs is the same as for the deterministic chain. However, the trajectories of the 
	quasiparticles are random walks, giving rise to diffusive entanglement growth. 
\end{abstract}
	
\maketitle

%############################################
\section{Introduction}
\label{sec:intro}

Understanding universality in out-of-equilibrium quantum many-body 
systems coupled to an environment remains a formidable challenge. 
In the regime of weak system-environment interactions and Markovianiaty, 
the average-state evolution can be described effectively by the Lindblad master 
equation~\cite{gorini1976completely,lindblad1976on,petruccione2002the}. 
While traditional studies focused primarily on average-state dynamics, 
recent attention has shifted toward understanding environment-induced 
fluctuations. 
One of the most urgent questions is whether a general framework 
to describe the statistics of these fluctuations 
exists~\cite{bernard2021can}. Indeed, unlike classical systems, for which a Macroscopic 
Fluctuation Theory ($MFT$) has been formulated~\cite{bertini2015macroscopic}, its generalization 
to the quantum many-body realm remains elusive. 
Such a framework must account for intrinsically quantum features, 
for instance, entanglement and its statistics. 
In \emph{isolated} one-dimensional integrable systems entanglement and 
local conserved quantities spread 
ballistically after quantum quenches~\cite{calabrese-2005,fagotti2008evolution,alba2017entanglement,alba2018entanglement,klobas2021exact,klobas2021entanglement}. Numerical evidence suggests ballistic 
entanglement spreading also in nonintegrable 
systems~\cite{kim2013ballistic}, despite diffusive energy transport.  
Ballistic spreading of entanglement in chaotic systems is further supported 
in random unitary circuits, and in a so-called ``membrane 
picture''~\cite{nahum2017quantum,jonay2018coarse}. Violations of 
this ballistic spreading paradigm are observed in some random unitary 
circuits~\cite{znidaric2020entanglement} (see~\cite{morralyepes2025disentangling} for a recent 
result). For \emph{open} quantum systems, 
the average-state  (e.g., obtained via Lindblad dynamics) 
typically saturates to a featureless, i.e., classical-like,  
steady state. Yet, in quadratic fermionic and bosonic systems~\cite{prosen2008third} 
the intermediate-time entanglement dynamics exhibits rich behavior, reminiscent of  
the quasiparticle picture in the unitary 
evolution~\cite{alba2021spreading,carollo2022dissipative,alba2022hydrodynamics,alba2023logarithmic,caceffo2023entanglement,caceffo2024faten}. Recently, there is mounting evidence that the 
statistics of entanglement fluctuations may harbour  
deeper universal signatures, such as entanglement-driven phase 
transitions~\cite{skinner2019measurement,jian202measurement},

Given this context, a paradigmatic model for studying fluctuating 
diffusive quantum dynamics is highly desirable. Such a model could serve 
as a playground for developing and validating effective descriptions, much 
like the classical symmetric exclusion process ($SEP$) played a pivotal 
role in establishing Macroscopic Fluctuation Theory ($MFT$)~\cite{bernard2021can}. 
Very recently, the so-called Quantum Symmetric Simple Exclusion Process ($QSSEP$) emerged as a promising 
candidate~\cite{bauer2019stochastic,bauer2019equilibrium,bernard2019open,bauer2020universal,bernard2021solution,hruza2023coherent,barraquand2025introduction}. The model describes a system of freely-hopping fermions on 
a ring with noisy, in both space and time, hopping amplitudes. For each realization of the 
noise the model is free-fermionic. 
The dynamics of the average state reduces to that of the classical Symmetric Simple Exclusion 
Process~\cite{mallick2015the} ($SSEP$), 
whereas fluctuations are beyond the classical regime. 
Interestingly, the full statistics of the  steady-state fluctuations 
can be determined~\cite{barraquand2025introduction}, thanks to nontrivial topological 
constraints on the noise-averaged moments of the correlators. 
Extensions of the $QSSEP$ include an asymmetric variant~\cite{jin2020from,bernard2022dynamics}, 
and a boundary-driven bosonic version~\cite{bernard2025large}. 
The probability distribution of the steady-state entanglement 
after a quantum quench from a product state was derived in Ref.~\cite{bernard2021entanglement} by 
using Random Matrix Theory (see also Ref.~\cite{bernard2021entanglement}). 
Yet, the full-time entanglement dynamics remains an open problem.

%
%############################################
\begin{figure}[t]
\centering
\includegraphics[width=.9\linewidth]{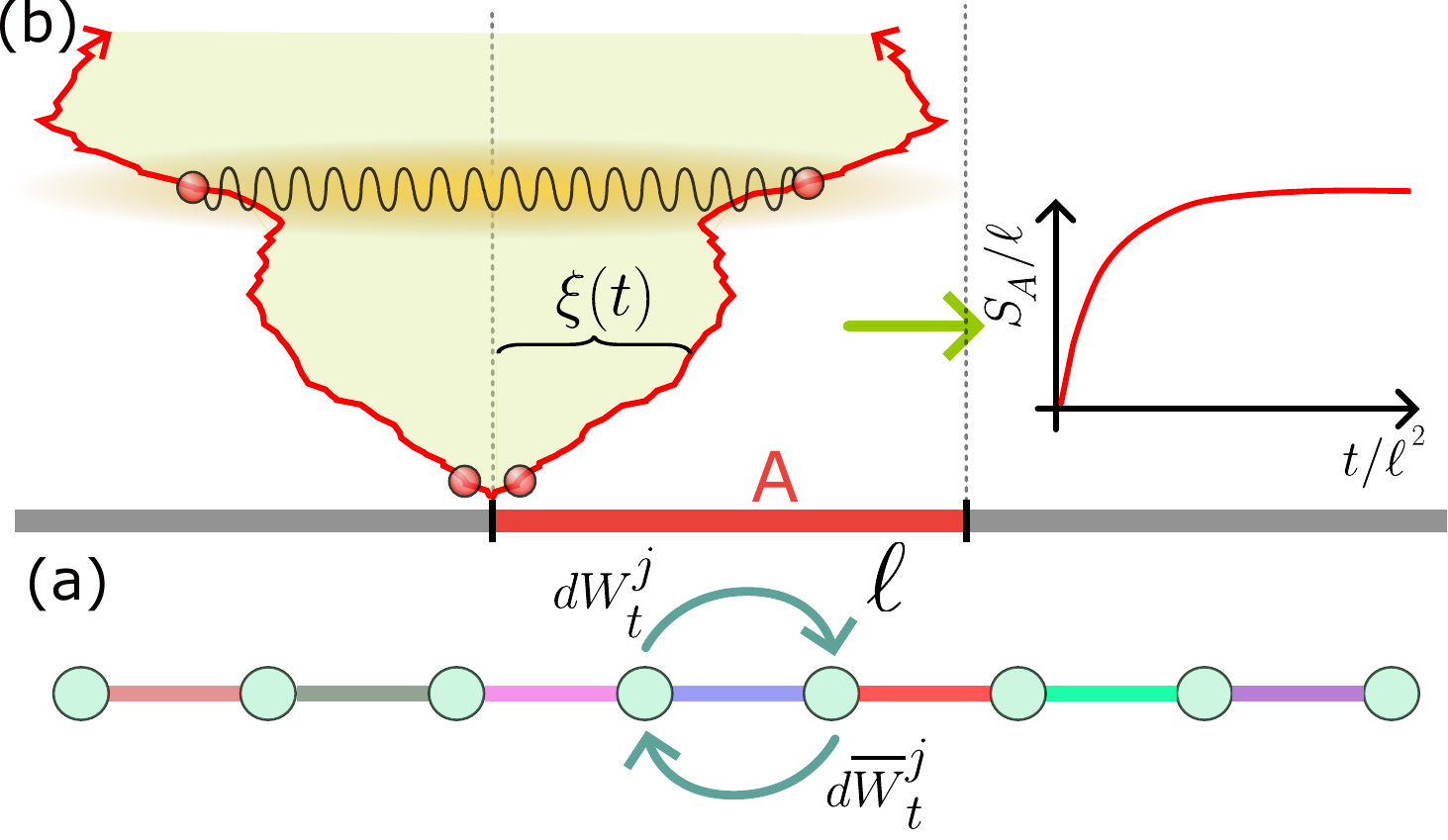}
\caption{ Definition of the $\nu$-$QSSEP$ model (a). A system of noninteracting 
	fermions evolves under the infinitesimal hopping 
	Hamiltonian $dH_t$ (cf.~\eqref{eq:ham}) with space-time stochastic hopping amplitudes 
	$dW^j_t,d\overline{W}_t^j$, the bar denoting complex conjugation.  
	Here $dW_{t}^j$  are random variables \emph{\`a la} It\^o, 
	which depend on time and on the position $j$. In the $\nu$-$QSSEP$ we 
	have $dW_{t}^j=dW_{t}^{j'}$ if $j=j'$ $\mathrm{mod}\,\nu$, with $\nu>1$ 
	an integer. (b) Quasiparticle picture for entanglement spreading after a quench 
	for $\nu=1$, i.e., spatially uniform noise. Pairs of entangled quasiparticles 
	are produced after the quench. The entanglement content of the pair is the same 
	as in the deterministic chain. The trajectories of the quasiparticles 
	forming an entangled pair are  mirror-symmetric. All the pairs undergo the same dynamics.  
	The entanglement between a subsystem $A$ of size $\ell$ 
	is proportional to $\min(2|\xi|,\ell)$, which is the number of pairs that are 
	shared between $A$ and its complement. 
	The distance $\xi(t)$ travelled by the quasiparticles 
	is stochastic and distributed with $P(\xi,t)=e^{-\xi^2/(4\gamma t)}/
	(2\sqrt{\pi\gamma t})$, giving rise to diffusive entanglement spreading. 
}
\label{fig:intro}
\end{figure} 
%############################################
%

Here we investigate out-of-equilibrium entanglement dynamics in a generalized 
$QSSEP$ model, in which we partially restore space-translation invariance, 
while retaining the randomness in time of the hopping amplitudes. 
Precisely, at each time step the hopping amplitudes are defined in a unit 
cell of size $\nu$, which is then repeated over the chain. The case with $\nu=1$ 
corresponds to spatially uniform noise. For $\nu=2$ one has different noise 
on even and odd sites. In the limit $\nu\to\infty$ one recovers the $QSSEP$. 
Our $\nu$-$QSSEP$ model is illustrated in Fig.~\ref{fig:intro} (a). 
Similar to the $QSSEP$,  the model is Gaussian for each realization of the noisy amplitudes. 
Hence, full information about the dynamics is encoded in the two-point fermionic 
correlation matrix. 
The stochastic differential equations governing the noise-averaged  moments of the correlations close on 
themselves, similar to the $QSSEP$, and in contrast with other noisy fermionic models, such as 
fermions with density monitoring~\cite{cao2019entanglement} (see also Ref.~\cite{carollo2022entangled}). We derive these equations for 
arbitrary $\nu$. We discuss in detail the case with $\nu=1$ and $\nu=2$. 
We show that the moments of the steady-state correlators 
satisfy non trivial identities, and can be obtained as the 
Haar average with random momentum-dependent  $SU(\nu)$ matrices. 

Now, while introducing a spatially modulated noise might seem a bit artificial, there are 
several reasons why this is interesting. The main one is that unlike the original QSSEP, 
it is possible to obtain exact results for the intermediate time dynamics in the 
$\nu$-QSSEP. Moreover, despite the structured noise the dynamics retains the main features of 
the QSSEP. Precisely, the dynamics of local observables and of the entanglement entropy is diffusive. 
As we are going to show, already for $\nu\sim 5$ the dynamics is numerically indistinguishable from that in the QSSEP. Furthermore, the study of quantum dynamics in the $\nu$-QSSEP can help unveil universal features. For instance, although the dynamics is diffusive for any $\nu$, it exhibits quite nontrivial correlations patterns. Precisely, for $\nu=1$ the dynamics exhibits only pairwise correlations, which underlies the straightforward applicability of the quasiparticle picture to describe entanglement dynamics. On the other hand, already for $\nu=2$ correlations between more than two quasiparticles develop, and they can be described analytically~\cite{alba2026work}. It is likely that such multiparticles correlations are present also in the QSSEP, although it is difficult to characterize them analytically.

For $\nu=1$ we show that it is possible to characterize analytically the 
full-time dynamics of the noise-averaged entanglement entropy, and in general of any 
sufficiently well-behaved function of the reduced density matrix of a subsystem 
$A$ (see Fig.~\ref{fig:intro}). The dynamics of 
entanglement-related quantities is understood in terms of a stochastic generalization of the 
quasiparticle picture~\cite{alba2021generalized}. 
Salient features of the quasiparticle picture are illustrated in Fig.~\ref{fig:intro} (b). 
Similar to the deterministic model, entanglement is transported by pairs 
of correlated quasiparticles. Remarkably, the 
entanglement entropy shared by the members of the pair is the same as in the 
deterministic model. The trajectories of the quasiparticles are stochastic. At the leading order in 
the limit $t,\ell\to\infty$, with $t/\ell^2$ fixed, and for each instance of the dynamics 
all the entangled pairs produced in the chain undergo the same random walk. Moreover, 
the trajectories of the quasiparticles forming an entangled pair are mirror-symmetric 
with respect to the emission point (see Fig.~\ref{fig:intro} (b)). 
Thus, the entanglement dynamics depends only on the distance $\xi(t)$ from the emission point. 
The entanglement between a subregion $A$ and the rest is proportional to the number of entangled pairs 
shared between $A$ and its complement. Since for each noise realization all the entangled pairs 
perform the same random walk, the number of shared pairs is $\min(2|\xi|,\ell)$, which is 
reminiscent of the result in the deterministic model. 
After averaging over the noise realizations, one obtains the entanglement entropies 
as 
\begin{equation}
	\label{eq:ent-intro}
	\langle S_n\rangle =\int_{-\infty}^\infty \!\!\!\!\!d\xi \min(2|\xi|,\ell)s_n P(\xi,t). 
\end{equation}
As anticipated, $s_n$, with $n$ an integer, are the \emph{thermodynamic} R\'enyi 
entropies  of the Generalized Gibbs Ensemble 
($GGE$)  that describes the steady state after the quench in the deterministic 
model~\cite{calabrese2016introduction,vidmar2016generalized}. In~\eqref{eq:ent-intro}, 
$P(\xi,t)$ is the heat kernel 
$P(\xi,t)=1/(2\sqrt{\pi\gamma t})e^{-\xi^2/(4\gamma t)}$, with $\gamma$ 
the square strength of the hopping amplitudes. Eq.~\eqref{eq:ent-intro} 
implies that entanglement spreading is diffusive, in contrast with the ballistic 
dynamics in the deterministic chain. The stochastic quasiparticle picture~\eqref{eq:ent-intro} holds in the 
limit $t,\ell\to\infty$ with $t/\ell^2$ fixed. Finally, we numerically verify that 
the diffusive entanglement scaling holds true for generic $\nu$, even in the 
limit $\nu\to\infty$. Eq.~\eqref{eq:ent-intro} captures qualitatively the entanglement spreading 
for generic $\nu$, although it is exact  only at $\nu=1$. On the other 
hand, we numerically show that the entanglement dynamics for $\nu=5$ is quite close to that 
at $\nu\to\infty$, at least for the quench from the N\'eel state. 
Eq.~\eqref{eq:ent-intro} is, at least to the best of our knowledge the first 
generalization of the quasiparticle picture to stochastic quantum dynamics.

The manuscript is structured as follows. In Section~\ref{sec:qssep-def} we introduce the 
$\nu$-$QSSEP$ model. In Section~\ref{sec:f-corr} we derive the equations describing the 
dynamics of the moments of fermionic correlation functions in momentum space, which contain 
full information on the statistics of observables. In particular, in Section~\ref{sec:nu-1} 
we focus on the $\nu$-$QSSEP$ with $\nu=1$. In Section~\ref{sec:nu-2} we 
discuss the moments of the correlators for $\nu=2$. In Section~\ref{sec:nu-2-steady} we show that the 
steady-state moments satisfy nontrivial relationships. In Section~\ref{sec:cons} 
we discuss the relationship between  the steady-state values of the moments and 
the initial-state expectation values of conserved quantities. In Section~\ref{sec:symm} 
we discuss the Haar invariance of steady-state observables,  focussing on 
the case with $\nu=2$. 
In Section~\ref{sec:e-dyn} we derive the stochastic quasiparticle 
picture for the entanglement dynamics. In Section~\ref{sec:numerics} we discuss numerical results supporting the 
results of Section~\ref{sec:e-dyn}, for the moments of the fermionic 
correlator (in Section~\ref{sec:num-1}) and for the von Neumann entropy (in Section~\ref{sec:num-2}). 
We conclude and discuss future directions in Section~\ref{sec:concl}. In Appendix~\ref{app:1} we 
provide some details on the derivation of the quasiparticle picture presented in Section~\ref{sec:e-dyn}.

%############################################
\section{$\nu$-QSSEP: Definition of the model} 
\label{sec:qssep-def}

Here we consider out-of-equilibrium dynamics in a generalization of the 
so-called $QSSEP$ model, which is defined by the infinitesimal Hamiltonian 
$dH_t$ defined as 
\begin{equation}
	\label{eq:ham}
	dH_t=\sqrt{\gamma}\sum_{j=1}^L\left(c_{j+1}^\dagger c_jdW_t^j+c^\dagger_j 
	c_{j+1}d\overline{W}_t^j\right)
\end{equation}
where $dW_{t}^j$ are random variables \emph{\`a la} It\^o, $\gamma$ a coupling constant, 
and $L$ the length of the chain. We employ periodic boundary conditions. 
The horizontal bar in $d\overline{W}_t^j$ denotes complex conjugation. 
Eq.~\eqref{eq:ham} describes a system of free fermions hopping 
in a chain with hopping amplitudes that are Brownian variables in both space and time. 
From~\eqref{eq:ham}, the infinitesimal Hamiltonian matrix $dh_t$ is 
\begin{equation} 
\label{eq:ham-mat} 
[dh_t]_{nm}=\sqrt{\gamma}\left(dW_t^n
\delta_{n-m,1}+d\overline{W}_t^m \delta_{m-n,1}\right).  
\end{equation}
We consider dynamics from an initial state $|\psi_0\rangle$ (quantum quenche) that is not 
eigenstate of~\eqref{eq:ham}. The time-evolved wavefunction satisfies the equation as 
$|\psi_{t+dt}\rangle=e^{-idH_t}|\psi_t\rangle$. Here we restrict ourselves to 
Gaussian initial states, i.e., satisfying Wick's theorem. It is straightforward to 
show that the dynamics under~\eqref{eq:ham} preserves Gaussianity. 
From the evolution of the wavefunction, we obtain the 
density matrix $\rho_t$ as
\begin{equation}
	\label{eq:rho}
\rho_t=|\psi_t\rangle\langle\psi_t|
\end{equation}
In the original $QSSEP$ model the amplitudes 
$dW_t^j$ are independent from site to site. Here we define the $\nu$-$QSSEP$  
by considering spatially-modulated random amplitudes. 
Precisely, we require that $W_t^n=W_t^{m}$ for 
$n=m\,\mathrm{mod}\,\nu$. For $\nu\ge L$, one recovers the standard $QSSEP$. 
For $\nu=1$ the noise $dW_t^j$ is the same on all sites, 
but random in time. For generic $\nu$ the noise is defined 
in a $\nu$-site unit cell. Our setup is depicted in Fig.~\ref{fig:intro} (a). 
The hopping amplitudes satisfy the standard It\^o relationships as 
\begin{equation} 
	\label{eq:ito-rule}
	dW_t^n d\overline{W}^m_t=\delta_{n-m,0\,\mathrm{mod}\,\nu},\quad 
	\langle dW_t^n\rangle_\mathrm{noise}=0.
\end{equation}
The evolution of the noise-averaged density matrix (cf.~\eqref{eq:rho}) 
is straightforwardly obtained by using~\eqref{eq:ito-rule} as 
\begin{multline}
\label{eq:lind}
\langle\rho_{t+dt}\rangle=\langle e^{idH_t}\rho_t e^{-idH_t}\rangle
=\langle\rho_t\rangle\\+
\gamma\!\!\!\sum_{\substack{j=j'\!\!\!\!\mod\nu\\\alpha=l,r}}
\!\!\Big(L_{\alpha,j}\langle\rho_t\rangle L_{\alpha,j'}^\dagger 
-\frac{1}{2}\Big\{L_{\alpha,j}L_{\alpha,j'}^\dagger ,\langle\rho_t\rangle\Big\}\Big)
\end{multline}
where we defined the Lindblad operators 
\begin{equation}
	\label{eq:lin-op}
	L_{l,j}=L_{r,j}^\dagger=c_{j+1}^\dagger c_j. 
\end{equation}
Eq.~\eqref{eq:lind} with~\eqref{eq:lin-op} for $\nu\to\infty$ reduces to the standard 
Lindblad equation describing a system subjected to incoherent nearest-neighbour hopping. 
This is similar to the setup investigated in Ref.~\cite{eisler2011crossover}, 
although  the standard Heisenberg term $[dH_t,\rho_t]$ is absent in~\eqref{eq:lind} 
due to the noise average. For finite $\nu$ the Lindblad operators~\eqref{eq:lin-op} 
become nonlocal in space since $j,j'$ can be different in~\eqref{eq:lind}. 

In the following we provide exact results for the steady-state value of 
observables after dynamics starting from arbitrary initial states. 
Moreover, for some low-entangled initial states we provide the 
full-time dynamics of entanglement-related quantities. 
Precisely, we consider dynamics starting from the fermionic N\'eel 
state $|N\rangle$ and the dimer state $|D\rangle$ defined as 
\begin{align}
	\label{eq:neel}
	& |N\rangle=\prod_{j=1}^{L/2}c^\dagger_{2j-1}|0\rangle\\
	\label{eq:dimer}
	& |D\rangle=\prod_{j=1}^{L/2}(c^\dagger_{2j-1}+c_{2j}^\dagger)/\sqrt{2}|0\rangle, 
\end{align}
where $|0\rangle$ is the fermionic vacuum state. 

%############################################
\subsection{Dynamics of correlation functions}
\label{sec:f-corr}

Here we are interested in the fermionic two-point correlation function
$G_{nm}$ defined as 
\begin{equation}
	\label{eq:gnm}
	G_{nm}=\langle\psi_t |c^\dagger_n c_m|\psi_t\rangle, 
\end{equation}
Since the dynamics under~\eqref{eq:ham} preserves the Gaussianity of
the initial state, $G_{nm}$ contains full information about the time-dependent 
state $\rho_t$ of the system.  The time-dependent correlation function  $G$  
is obtained from the evolved state $|\psi_t\rangle$ as 
\begin{equation} 
	\label{eq:G-te}
	G(t+dt)= e^{-idh_t}G(t) e^{idh_t}, 
\end{equation}
where $dh_t$ is given in~\eqref{eq:ham-mat}. 
Equivalently, by expanding the exponentials in~\eqref{eq:G-te}, 
up to second order one obtains 
\begin{multline} 
\label{eq:G1-ev} dG_{nm}=\gamma \left[\Delta^\mathrm{dis}
G\right]_{nm}dt -i\sqrt{\gamma}\big(dW_t^n G_{n-1,m}+\\
d\overline{W}_t^{n+1}
G_{n+1,m} -d\overline{W}_t^m G_{n,m-1}-dW_t^{m+1}G_{n,m+1}\big). 
\end{multline}
Formally, Eq.~\eqref{eq:G1-ev} is the same as in the $QSSEP$~\cite{bauer2019equilibrium}. 
However, our definition of $\Delta^\mathrm{dis} G$ is different, and it is 
given as 
\begin{equation} 
	\label{eq:D1}
		[\Delta^\mathrm{dis} G]_{nm}:=G_{n-1,m-1}-2G_{nm}+G_{n+1,m+1} 
\end{equation}
for $n=m\,\mathrm{mod}\,\nu$,  and 
\begin{equation}
	\label{eq:D2}
	[\Delta^\mathrm{dis}G]_{nm}:=-2G_{nm}, 
\end{equation}
for $n\ne m\,\mathrm{mod}\,\nu$.
Now, after averaging over the noise $dW_t^n$ in Eq.~\eqref{eq:G1-ev}, only the 
first row survives. 
For the $QSSEP$, i.e., in the limit $\nu\to\infty$, only the diagonal terms 
with $n=m$ in~\eqref{eq:D1} are present, whereas for $n\ne m$ 
Eq.~\eqref{eq:D2} remains the same. Instead, for the $\nu$-$QSSEP$ 
Eq.~\eqref{eq:G1-ev} together with~\eqref{eq:D2} 
imply that $\langle G_{nm}\rangle_\mathrm{noise}$ 
vanishes exponentially with time for $n-m\ne0\,\mathrm{mod}\,\nu$, 
whereas it attains a nonzero steady-state value otherwise. 

Here we are interested in the dynamics of arbitrary functions of $G_{nm}$. 
To understand that, it is crucial to derive the dynamics of the noise-averaged 
moments of $G_{nm}$, such as  the second moment $\langle G_{n_1m_1}G_{n_2m_2}\rangle$. 
By applying the standard It\^o rule, 
this is obtained as $d\langle G_{n_1m_1}G_{n_2m_2}\rangle=
\langle dG_{n_1m_1}G_{n_2m_2}+G_{n_1m_1 }dG_{n_2m_2}+
dG_{n_1m_1}dG_{n_2m_2}\rangle$, where one has to 
use~\eqref{eq:G1-ev} to compute each of the three terms. 
Notice that third term is nonzero solely due to the noise. 
One obtains 
\begin{multline} \label{eq:G2-ev} \gamma^{-1}\frac{d}{dt} \langle G_{n_1m_1}G_{n_2m_2}\rangle=
	\left[\left\langle\Delta^\mathrm{dis} G \right\rangle
	\right]_{n_1m_1} G_{n_2m_2} \\+
	G_{n_1m_1}\left[\left\langle\Delta^\mathrm{dis}G\right\rangle
	\right]_{n_2m_2} -\Big\langle
	\delta_{n_1\pm 1,n_2}G_{n_1\pm
		1,m_1}G_{n_2\mp1,m_2}\\
		-\delta_{n_1,m_2}G_{n_1\pm1,m_1}G_{n_2,m_2\pm1}
		-\delta_{m_1-n_2,0}G_{n_1,m_1\pm 1}G_{n_2\pm 1,m_2}\\+
		\delta_{m_1,m_2\mp1}G_{n_1,m_1\mp1}G_{n_2,m_2\pm1}
		\Big\rangle,
\end{multline}
where one has to sum over the $\pm$, and all the Kronecker delta 
functions are defined $\mathrm{mod}\,\nu$. In~\eqref{eq:G2-ev} 
$\Delta^\mathrm{dis} G$ is as defined in~\eqref{eq:D1}. 
By direct inspection of~\eqref{eq:G2-ev}, it is straightforward to show that 
in the steady state at $t\to\infty$, only the correlators $\langle G_{n n}G_{mm}\rangle$ 
and  $\langle G_{n m}G_{mn}\rangle$, where indices are equal $\mod\,\nu$, are  nonzero. 

To proceed, it is convenient to exploit the invariance of~\eqref{eq:ham} under translations 
by $\nu$ sites, going to momentum space. Thus, we  define the Fourier transformed 
correlator $G_{kq}^{\rho\sigma}$ as 
\begin{equation} \label{eq:ft} 
	G_{\nu n+\rho,\nu
m+\sigma}=\frac{\nu}{L}\sum_{k,q}e^{-i k n +iq m}G^{\rho\sigma}_{k,q},
\end{equation}
with $k,q=2\pi/(L/\nu)j$, with $j=0,1,\dots L/\nu-1$, and 
$\rho,\sigma=0,1,\dots,\nu-1$ the unit cell indices. 
At fixed $k,q$, $G_{kq}^{\rho\sigma}$ is a $\nu\times\nu$ matrix. 
In general $G_{kq}$ is not diagonal, except for dynamics starting from 
translational invariant initial states. 
We can perform the Fourier transform of  
Eq.~\eqref{eq:D1} and~\eqref{eq:D2}, obtaining for $\rho=\sigma$ 
\begin{multline}
\label{eq:Delta}
	\left[\Delta^\mathrm{dis}
	G\right]_{k,q}^{\rho\rho}=
	(
	e^{i(k-q)}\delta_{\rho=0}+\delta_{\rho\ne0})G_{kq}^{\rho-1\rho-1}\\+
(e^{-i(k-q)}\delta_{\rho=\nu-1}+\delta_{\rho\ne\nu-1})G_{kq}^{\rho+1\rho+1}
-2G_{kq}^{\rho\rho}, \end{multline}
where $\delta_{\rho=0}$ and $\delta_{\rho\ne0}$ enforce the conditions  
$\rho=0$ and $\rho\ne0$, respectively. For $\rho\ne \sigma$ we have 
\begin{equation} \label{eq:Delta-1}
\left[\Delta^\mathrm{dis}G\right]_{kq}^{\rho\sigma}=-2G_{kq}^{\rho\sigma}.
\end{equation}
Thus, the evolution of the correlation function~\eqref{eq:G1-ev} 
in momentum space takes the simple form as 
\begin{multline} \label{eq:Gk-evol} dG_{kq}^{\rho\sigma}=\gamma
	dt\left[\Delta^{\mathrm{dis}}G\right]_{kq}^{\rho\sigma}
	-i\sqrt{\gamma}\Big[
		dW_t^{\rho}(\delta_{\rho=0}e^{ik}+\delta_{\rho\ne0})G_{kq}^{\rho-1\sigma}\\
		+d\overline{W}_t^{\rho+1}(\delta_{\rho=\nu-1}e^{-ik}+\delta_{\rho\ne\nu-1})G_{kq}^{\rho+1\sigma}\\
		-d\overline{W}_t^{\sigma}(\delta_{\sigma=0}e^{-iq}+\delta_{\sigma\ne0})G_{kq}^{\rho\sigma-1}
		-dW_t^{\sigma+1}(\delta_{\sigma=\nu-1}e^{iq}\\
		+\delta_{\sigma\ne\nu-1})G_{kq}^{\rho\sigma+1}
\Big],  \end{multline}
where $[\Delta^\mathrm{dis}G]_{kq}^{\rho\sigma}$ is defined
in~\eqref{eq:Delta} and~\eqref{eq:Delta-1}, and the indices $\rho\pm1,\sigma\pm1$ are 
computed $\mathrm{mod}\,\nu$.  Now, for $\rho\ne\sigma$, after taking the noise average, 
Eq.~\eqref{eq:Gk-evol} gives the exponential decay as 
\begin{equation}
	\label{eq:nondiag}
	\frac{d\langle
G_{kq}^{\rho\sigma}\rangle}{dt}=-2\gamma\langle
G_{kq}^{\rho\sigma}\rangle, \end{equation} 
On the other hand, for $\rho=\sigma$, Eq.~\eqref{eq:Gk-evol} 
gives 
\begin{multline} 
	\label{eq:one-point}
	\frac{d\langle
G_{kq}^{\rho\rho}\rangle}{dt}=\gamma\langle
(e^{i(k-q)}\delta_{\rho=0}+\delta_{\rho\ne0})G_{kq}^{\rho-1\rho-1}\\
+(e^{-i(k-q)}\delta_{\rho=0}+\delta_{\rho\ne0})
G_{kq}^{\rho+1\rho+1}-2G_{kq}^{\rho\rho}\rangle. 
\end{multline}
It is interesting to investigate the steady-state value of $G_{kq}^{\rho\rho}$. 
To have a nonzero steady-state value, the right-hand side of~\eqref{eq:one-point} has to 
vanish. From~\eqref{eq:one-point}, one 
obtains a system of $\nu$ equations. It is straightforward to check that 
a nonzero steady-state value is possible only for 
$k=q$. Moreover, one obtains that $G_{kk}^{\rho\rho}$ does not depend on $\rho$. 

Similar to the $QSSEP$ model~\cite{bauer2019equilibrium}, 
the steady-state value of the correlator and its moments can be 
obtained from the expectation value on the initial state of conserved quantities. 
For instance, by using the evolution equation~\eqref{eq:Gk-evol}, it is straightforward to check  
that $N_k$ defined as 
\begin{equation} 
	\label{eq:Nk}
N_k=\sum_{\rho=0}^{\nu-1} G_{kk}^{\rho\rho} \end{equation}
is conserved during the dynamics for any $k$, even for a single realization of 
the noise. This implies that in the steady state one has 
\begin{equation}
	\label{eq:G-ss}
	\langle G_{kq}^{\rho\sigma}\rangle=\frac{N_k}{\nu}\delta_{kq}
	\delta_{\rho\sigma},  
\end{equation}
The result~\eqref{eq:G-ss} holds for any value of $\nu$. The fact that 
the nonzero steady-state values of the correlators can be determined from 
the conserved quantities remains true for higher moments of the correlation functions, 
and is reminiscent of a similar behavior in the $QSSEP$~\cite{bauer2019equilibrium}. 

The building block to obtain the dynamics of the noise-averaged moments of 
$G^{\rho\sigma}_{kq}$ is the dynamics of the four-point function  
$G_{k_1,q_1}^{\rho_1\sigma_1}G_{k_2,q_2}^{\rho_2\sigma_2}$. By 
using~\eqref{eq:Gk-evol}, we obtain 
\begin{multline} \label{eq:evol-square} d(G_{k_1,q_1}^{\rho_1\sigma_1}
	G_{k_2,q_2}^{\rho_2\sigma_2})=
	(dG_{k_1,q_1}^{\rho_1\sigma_1})G_{k_2,q_2}^{\rho_2\sigma_2}
	\\
	+G_{k_1,q_1}^{\rho_1\sigma_1}(dG_{k_2,q_2}^{\rho_2\sigma_2})-
	\gamma K_{12}dt 
\end{multline}
where we defined $K_{ij}$ as 
\begin{multline}
	\label{eq:K-Ito}
	K_{ij}=-\delta_{\rho_i,\sigma_j}\widetilde{G}_{k_i,q_i}^{\rho_i-1\sigma_i}
		\widetilde{G}_{k_j,q_j}^{\rho_j\sigma_j-1}
		\\+\delta_{\rho_i,\rho_j+1}
		\widetilde{G}_{k_i,q_i}^{\rho_i-1\sigma_i}
		\widetilde{G}_{k_j,q_j}^{\rho_j+1\sigma_j}
		+\delta_{\rho_i+1,\rho_j}\widetilde{G}_{k_i,q_i}^{^{\rho_i+1\sigma_i}}
		\widetilde{G}_{k_j,q_j}^{\rho_j-1\sigma_j}\\
		-\delta_{\rho_i,\sigma_j}
		\widetilde{G}_{k_i,q_i}^{
		\rho_i+1\sigma_i}\widetilde{G}_{k_j,q_j}^{\rho_j\sigma_j+1}
		-\delta_{\sigma_i,\rho_j}\widetilde{G}_{k_i,q_i}^{\rho_i\sigma_i-1}
		\widetilde{G}_{k_j,q_j}^{\rho_j-1\sigma_j}\\
		+\delta_{\sigma_i,\sigma_j+1}
		\widetilde{G}_{k_i,q_i}^{\rho_i\sigma_i-1}
		\widetilde{G}_{k_j,q_j}^{\rho_j\sigma_j+1}
		-\delta_{\sigma_i,\rho_j}
		\widetilde{G}_{k_i,q_i}^{\rho_i\sigma_i+1}
		\widetilde{G}_{k_j,q_j}^{\rho_j+1\sigma_j}\\
		+\delta_{\sigma_i+1,\sigma_j}
		\widetilde{G}_{k_i,q_i}^{\rho_i\sigma_i+1}
		\widetilde{G}_{k_j,q_j}^{\rho_j\sigma_j-1}, 
\end{multline}
with $\widetilde{G}^{\rho\sigma}_{kq}$ defined as 
\begin{align}
	&\widetilde{G}_{k,q}^{\rho-1,\sigma}=(\delta_{\rho=0}e^{ik}+\delta_{\rho\ne0})G_{k,q}^{\rho-1\sigma}\\
	&\widetilde{G}_{k,q}^{\rho+1,\sigma}=(\delta_{\rho=\nu-1}e^{-ik}+\delta_{\rho\ne\nu-1})G_{k,q}^{\rho+1\sigma}\\
	&\widetilde{G}_{k,q}^{\rho,\sigma-1}=(\delta_{\sigma=0}e^{-iq}+\delta_{\sigma\ne0})G_{k,q}^{\rho\sigma-1}\\
	&\widetilde{G}_{k,q}^{\rho,\sigma+1}=(\delta_{\sigma=\nu-1}e^{iq}+\delta_{\sigma\ne\nu-1})G_{k,q}^{\rho\sigma+1}. 
\end{align}
Again, the indices $\rho_j,\sigma_j$ are defined $\mathrm{mod}\,\,\nu$. 
The term $K_{12}$ in~\eqref{eq:evol-square} originates from 
the application of the It\^o  rule to 
$dG_{k_1q_1}^{\rho_1\sigma_1}dG_{k_2q_2}^{\rho_2\sigma_2}$. 
The system of $\nu$ equations in~\eqref{eq:evol-square} fully determines the 
dynamics of the product $G_{k_1q_1}G_{k_2q_2}$. Eq.~\eqref{eq:evol-square} 
is straightforwardly generalized to the higher moments of $G_{kq}$. 
Indeed, for the generic $n$-th moment, we have that 
$d(G^{\rho_1\sigma_1}_{k_1q_1}G^{\rho_2\sigma_2}_{k_2q_2}
\cdots G^{\rho_n\sigma_n}_{k_nq_n})$ contains all 
the terms with a single $dG^{\rho_j\sigma_j}_{k_jq_j}$, as well as all the It\^o 
terms as $dG_{k_jq_j}^{\rho_j\sigma_j}dG_{k_lq_l}^{\rho_l\sigma_l}$ with $j\ne l$. 
The former give rise to contributions of the form~\eqref{eq:G1-ev}, whereas the latter give 
$\gamma K_{jl}dt$, with $K_{jl}$ as in~\eqref{eq:K-Ito}.

%############################################
\subsection{Full-time dynamics for $\nu=1$} 
\label{sec:nu-1}

Before proceeding let us discuss the case with $\nu=1$, which corresponds to the 
situation in which the hopping amplitudes $dW_t^j$ do not depend on $j$, although they 
are random in time. Now, for any $k,q$, $G_{kq}$ is a number, and Eq.~\eqref{eq:Gk-evol}  becomes 
\begin{multline}
\label{eq:dyn}
dG_{kq}=\Big\{2\gamma[\cos(k-q)-1] dt
-i\sqrt{\gamma}(e^{ik}-e^{iq})dW_t\\-i\sqrt{\gamma}
(e^{-ik}-e^{-iq})d\overline{W}_t\Big\}G_{kq}, 
\end{multline}
from where it is clear that $G_{kk}$ is conserved by the dynamics for any 
$k$, even before the noise average. In contrast with generic $\nu$, 
for $\nu=1$ it is straightforward to obtain the full-time dynamics of arbitrary 
noise-averaged moments of the correlators. 
For instance, for $G_{kq}$, by taking the average over $d W_t,d\overline{W}_t$ in~\eqref{eq:dyn} 
one obtains the equation for $\langle G_{kq}\rangle$, whose solution  is $\langle G_{kq}
\rangle=G_{kq}^{\scriptscriptstyle(0)}e^{\varepsilon t}$, with $G_{kq}^{\scriptscriptstyle(0)}$ the initial condition 
and the ``energy'' $\varepsilon=2\gamma [\cos(k-q)-1]$. $\langle G_{kq}\rangle$ decays 
exponentially with time except for $k=q$. Let us consider the dynamics of 
$\langle G_{k_1q_1}G_{k_2q_2}\rangle$. From~\eqref{eq:evol-square} we obtain 
\begin{multline}
	\label{eq:four-nu-1}
	d\langle G_{k_1q_1}G_{k_2q_2}\rangle=2\gamma[-2-\cos(k_1-k_2)\\
		+\cos(k_1-q_1)+\cos(k_2-q_1)+\cos(k_1-q_2)\\
	+\cos(k_2-q_2)-\cos(q_1-q_2)]\langle G_{k_1q_1}G_{k_2q_2}\rangle dt
\end{multline}
Eq.~\eqref{eq:four-nu-1} implies that $\langle G_{k_1q_1}G_{k_2q_2}\rangle$ 
decays exponentially with time, except for $G_{k_1k_1}G_{k_2k_2}$ and 
$G_{k_1q_1}G_{q_1k_1}$. It is useful to introduce a pictorial notation for 
$G_{kq}$. Let us denote $G_{kq}$ with a directed arrow as 
\begin{equation}
	G_{kq}=\hspace{2pt}\graphlinefin{.35}{},
\end{equation}
where the left and right endpoints denote the momenta $k$ and $q$, respectively. 
It is easy to check that the quasimomenta configurations that correspond to 
correlators that have a nonzero steady state value are given by the diagrams 
that at each vertex have the same number of ingoing and outgoing lines, which 
implies momentum conservation at each vertex. For instance, the only nonzero 
correlators $\langle G_{k_1q_1}G_{k_2q_2}\rangle$ correspond to the diagrams 
\begin{equation}
\graphN{.5}\hspace{2pt}\graphN{.5},\hspace{4pt}
\graphFdeux{.5},\hspace{4pt}\graphNdeux{.5}. 
\end{equation}
Let us now observe that since Eq.~\eqref{eq:four-nu-1} can be generalized 
to arbitrary moments of the correlators $G_{kq}$, one can obtain the full-time 
dynamics of arbitrary observables. However, the generalization of~\eqref{eq:four-nu-1} 
become rapidly quite cumbersome. Still, as it will be clear in Section~\ref{sec:e-dyn}, 
for dynamics starting from translation-invariant initial states and in the long time 
limit only some diagrams contribute, which allows to determine the full-time entanglement 
dynamics. 

%############################################
\section{Dynamics in the $\nu$-$QSSEP$ with $\nu=2$}
\label{sec:nu-2}

Having discussed the case with $\nu=1$, here we focus on the $\nu$-$QSSEP$ with $\nu=2$. 
Precisely, we analyze the evolution equations~\eqref{eq:Gk-evol}, showing that the 
moments of the steady-state correlation functions satisfy a set of ``topological'' 
constraints, which allow to determine their value from initial-state data. 

Let us first consider the two-point function $\langle G_{kq}^{\rho\sigma}\rangle$, 
where $\rho,\sigma=0,1$. Eq.~\eqref{eq:Gk-evol} gives 
\begin{align}
	\label{eq:t-1}
	& d\langle G_{kq}^{00}\rangle=-2\gamma \langle G_{kq}^{00}\rangle+ 
	\gamma (1+e^{i(k-q)})\langle G_{kq}^{11}\rangle d t\\
	\label{eq:t-2}
	& d\langle G_{kq}^{11}\rangle=-2\gamma \langle G_{kq}^{11}\rangle+ 
	\gamma (1+e^{-i(k-q)})\langle G_{kq}^{00}\rangle d t. 
\end{align}
Clearly, Eq.~\eqref{eq:t-1} and~\eqref{eq:t-2} imply that $G_{kq}^{\rho\rho}$ 
is nonzero in the steady-state only for $k=q$, as anticipated in Section~\ref{sec:f-corr} 
for generic $\nu$. 
From~\eqref{eq:nondiag} one has that $\langle G_{kq}^{\rho\sigma}\rangle$ vanishes 
exponentially for $\rho\ne \sigma$. By using~\eqref{eq:t-1},~\eqref{eq:t-2}, 
and~\eqref{eq:nondiag} one can obtain the exact full-time dynamics of $G_{kq}^{\rho\sigma}$. 
This is not straightforward for more complicated moments of the correlation functions. 
For instance, let us focus on the second moment  $G_{k_1q_1}G_{k_2q_2}$. 

A direct analysis of~\eqref{eq:evol-square} shows that the only 
moments attaining nonzero steady-state values correspond to  
$\{\rho_1,\sigma_1,\rho_2,\sigma_2\}=\{0,0,0,0\},\{0,0,1,1,\},\{0,1,1,0\}$, together 
with those obtained by swapping $1\leftrightarrow 0$. 
The dynamics of the noise-averaged moment 
$\langle G_{k_1q_1}^{\rho_1\sigma_1}G_{k_2q_2}^{\rho_2\sigma_2}
\rangle$ is governed by a $6\times 6$ matrix $M$ as 
\begin{equation}
	\label{eq:gener}
	\frac{d}{dt}\langle G_{k_1q_1}G_{k_2q_2}\rangle=\gamma \sum_{\rho_j,\sigma_j}
	M_{\rho_1\sigma_1\rho_2\sigma_2}\langle G_{k_1q_1}^{\rho_1\sigma_1}
	G_{k_2 q_2}^{\rho_2\sigma_2}\rangle, 
\end{equation}
where the sum over $\rho_j,\sigma_j$ is restricted to the six configurations above. 
The matrix $M$ is constructed starting from any 
set of indices $\{\rho_1,\sigma_1,\rho_2,\sigma_2\}$ for which 
$G^{\rho_1\sigma_1}_{k_1q_1}G^{\rho_2\sigma_2}_{k_2q_2}$ is nonzero in the steady state, 
and iteratively applying~\eqref{eq:evol-square}. On obtains 
\begin{widetext}
\begin{equation} 
	\label{eq:M6}
	M=-5\mathds{1}_6+O_6+
	\begin{pmatrix} 0& e^{ik_2-iq_2}& e^{ik_2-iq_1}&
		e^{ik_1-iq_2} & e^{ik_1-iq_1} &-1\\ 
		e^{-ik_2+iq_2}&0&
		-2-e^{-iq_1+iq_2} & -2-e^{ik_1-ik_2}& -1& e^{ik_1-iq_1} \\
		e^{-ik_2+iq_1}& -2-e^{iq_1-iq_2}&0 & -1 & -2-e^{ik_1-ik_2} &
		e^{ik_1-iq_2} \\ 
		e^{-ik_1+iq_2}& -2-e^{-ik_1+ik_2}&-1& 0 &
		-2-e^{-iq_1+iq_2} & e^{ik_2-iq_1} \\ 
		e^{-ik_1+iq_1}& -1&
		-2-e^{-ik_1+ik_2}&  -2-e^{iq_1-iq_2}&0 & e^{ik_2-iq_2} \\ 
		-1&
		e^{-ik_1+iq_1}& e^{-ik_1+iq_2}&  e^{-ik_2+iq_1} &
		e^{-ik_2+iq_2} & 0 
\end{pmatrix}, 
\end{equation}
\end{widetext}
where $\mathds{1}_6$ is the identity matrix and $O_6$ is 
the matrix with all its entries equal to $1$. 
One can check that the eigenvalues of~\eqref{eq:M6} are nontrivial for generic $k_j,q_j$, 
implying that it is not straightforward to obtain the \emph{exact} finite-time dynamics. 
As anticipated, for some momenta configurations $k_j,q_j$, Eq.~\eqref{eq:M6} will 
have some zero eigenvalues, which correspond to nonzero steady-state values of 
$\langle G_{k_1q_1}^{\rho_1\sigma_1}G_{k_2q_2}^{\rho_2\sigma_2}\rangle$. 
For completeness, let us also consider 
$G_{k_1q_1}^{\rho_1\sigma_1}G_{k_2q_2}^{\rho_2\sigma_2}$, with 
$\{\rho_1,\sigma_1,\rho_2,\sigma_2\}=\{0,0,1,0\},\{1,0,0,0\},\{1,0,1,1\},\{1,1,1,0,\}$, 
i.e., having and odd number of $1$s. By employing~\eqref{eq:evol-square} one can check that 
this set of correlators is closed under the dynamics. 
Now the matrix governing the dynamics is $4\times 4$, and it reads as 
\begin{widetext}
\begin{equation}
	\label{eq:M4}
	M=-5\mathds{1}_4+1_4\otimes 1_4+
	\left(\begin{array}{cccc}
			 0& -2-e^{ik_1-ik_2} & e^{ik_1-iq_2} & e^{ik_1-iq_1}\\
			-2-e^{-ik_1+ik_2} &  0& e^{ik_2-iq_2} & e^{ik_2-iq_1}\\
			e^{-ik_1+iq_2} & e^{-ik_2+iq_2} & 0& -2-e^{-iq_1+iq_2}\\
			e^{-ik_1+iq_1} & e^{-ik_2+iq_1} & -2-e^{iq_1-iq_2} & 0
		\end{array}
	\right). 
\end{equation}
\end{widetext}
Although the eigenvalues of~\eqref{eq:M4} can be determined exactly, their expression is 
cumbersome, and we do not report it. However, one can check that all the eigenvalues 
of~\eqref{eq:M4} are negative for any choice of $k_j,q_j$. This implies that all the correlators 
vanish exponentially at long times. Finally, we should observe that the dynamics of 
$G^{01}_{k_1q_1}G^{00}_{kl_2q_2}$ is not accounted for in~\eqref{eq:M4}. However, it is easily 
obtained from $G^{11}_{k_1q_1}G^{10}_{k_2q_2}$ by swapping $1\leftrightarrow0$ and 
$(k_1,q_1)\leftrightarrow (k_2,q_2)$. In the next sections we show that although 
for $\nu=2$ it could be challenging to obtain the exact full-time dynamics of the generic 
moments of the correlators, their steady-state values satisfy some nontrivial constraints, 
which can be explained by Random Matrix Theory.

%############################################
\subsection{Steady-state correlators}
\label{sec:nu-2-steady}

To identify the moments $\langle G_{k_1q_1}^{\rho_1\sigma_1}
G_{k_2q_2}^{\rho_2\sigma_2}\rangle$  that attain nonzero steady-state values, one 
has to determine the kernel of $M_6$ in~\eqref{eq:M6}, i.e., the eigenvectors  
with zero eigenvalues, for all the configurations of the quasimomenta $k_j,q_j$. 
Now, one can straightforwardly 
check that Eq.~\eqref{eq:M6} has zero eigenvalues only if all the momenta $k_jq_j$ appear 
an even number of times.  To illustrate the correlators that are nonzero in the steady state 
we can employ the pictorial notation used for $\nu=1$ in Section~\ref{sec:nu-1}, ignoring 
the indices $\rho_j\sigma_j$ for now. 
The only quasimomenta configurations yielding nonzero steady-state 
results correspond to the diagrams 
\begin{equation} \label{eq:d3}
\graphN{.5}\hspace{2pt}\graphN{.5},\hspace{4pt}
\graphFdeux{.5},\hspace{4pt}\graphNdeux{.5},\hspace{4pt}\graphFdeuxr{.5}, 
\end{equation}
The terms in~\eqref{eq:d3} have different number of independent quasimomenta. 
Precisely, the two leftmost ones contain two indipendent
momenta, and they represent  $\langle G^{\rho_1\sigma_1}_{kk}
G_{qq}^{\rho_2\sigma_2}\rangle$ and
$\langle G_{kq}^{\rho_1\sigma_1}G_{qk}^{\rho_2\sigma_2}\rangle$, with $k\ne q$. 
The third term in~\eqref{eq:d3} corresponds to $\langle 
G^{\rho_1\sigma_1}_{kk}G^{\rho_2\sigma_2}_{kk}\rangle$, 
where a single momentum $k$ appears.  The last term is
$\langle G^{\rho_1\sigma_1}_{k q}G_{kq}^{\rho_2\sigma_2}\rangle $, with $k\ne q$. 
One can check that also for higher moments of $G_{kq}$ only the diagrams 
in which all the momenta appear an even number of times can yield nonzero steady-state 
values. Moreover, it is interesting to observe that all the  diagrams 
in~\eqref{eq:d3}, except the last one, satisfy the rule that at each vertex 
the total number of incoming and outgoing lines is the same, reflecting 
conservation of momentum at each vertex. For $\nu=1$ diagrams violating 
the momentum conservation rule were absent (see Section~\ref{sec:nu-1}). 

To fully determine which $\langle G^{\rho_1\sigma_1}_{k_1q_1}
G^{\rho_2\sigma_2}_{k_2q_2}\rangle$ are nonzero in the steady state, 
one has to determine for each of the diagrams in~\eqref{eq:d3} all 
the allowed values for the indices $\rho_j\sigma_j$. 
To proceed, let us first focus on the configurations satisfying momentum conservation at 
each vertex (the three leftmost diagrams in~\eqref{eq:d3}). By solving for the null space of~\eqref{eq:M6} for each of the 
diagrams in~\eqref{eq:d3}, one can check that the eigenvectors components are 
nonzero only if  at each vertex the number of incoming and outgoing $0$s and $1$s is the same. 
This means that vertices that satisfy momentum conservation also satisfy 
conservation of $0$ and $1$.  We verified that this rule generalizes to 
higher-point functions, and it is also valid for generic $\nu$. Moreover, 
further constraints on the indices $\rho_j,\sigma_j$ exist. 

To discuss them, it is convenient to introduce a pictorial notation, denoting  
with a directed red line the correlator $G_{kq}^{\rho\sigma}$ as 
$\graphlineintro{.25}{}{red}$, the endpoints being $\rho$ and $\sigma$. 
Now, for each diagram of~\eqref{eq:d3} there are several compatible red diagrams. 
The value of the red vertices is immaterial, the only constraint being 
that distinct vertices correspond to different values of $\rho,\sigma$. 
Moreover, for $\nu=2$ the 
allowed diagrams representing $\rho_j,\sigma_j$ can have at most two vertices, in contrast with 
the momentum diagrams (as, for instance, in~\eqref{eq:d3}), which can have an arbitrary number 
of vertices. Interestingly, diagrams corresponding  to nonzero 
steady-state moments of the correlator are not all independent but satisfy nontrivial relationships. 
These relationships are simpler if only vertices with only one ingoing and one outgoing 
momentum lines are present. For instance, the diagram $\graphN{.4}\hspace{4pt}\graphN{.4}$ gives 
\begin{equation}
	\label{eq:id4}
	\left\langle\hspace{2pt}\graphN{.5}\hspace{2pt}\graphN{.5}
	\hspace{3pt}\right\rangle_\mathrm{}\rightarrow  
	\langle\hspace{.5pt}\graphNdeuxred{.5}\hspace{.5pt}\rangle_\mathrm{}=
	\langle\hspace{2pt}\graphNred{.5}\hspace{2pt}\graphNred{.5}\hspace{3pt}\rangle_\mathrm{}, 
\end{equation}
where we restored the noise average $\langle\cdot\rangle$. 
Similarly, we obtain 
\begin{equation}
	\label{eq:id5}
	\langle\hspace{2pt}\graphFdeux{.5}\hspace{2pt}\rangle_\mathrm{}
	\rightarrow  \langle\graphNdeuxred{.5}\rangle_\mathrm{}=\langle\hspace{2pt}
	\graphFdeuxred{.5}
	\hspace{2pt}\rangle_\mathrm{}. 
\end{equation}
From~\eqref{eq:id4} and~\eqref{eq:id5} we conclude that all the moments that correspond to diagrams 
with vertices satisfying the conservation of momentum and of cell indices $0,1$ have the same value in the 
steady state. However, this does not remain true if there are vertices connecting 
more than two momentum lines. 
For instance, let us focus on the diagram with $k_1=q_1=k_2=q_2$. Now, from the null space of~\eqref{eq:M6} 
one obtains that 
\begin{equation}
	\label{eq:id6}
	\langle\graphNdeux{.5}\rangle_\mathrm{}\rightarrow  
	\langle\graphNdeuxred{.5}\rangle_\mathrm{}=
	\langle\hspace{1pt}\graphFdeuxred{.5}\hspace{1pt}\rangle_\mathrm{}+
	\langle\hspace{1pt}\graphNred{.5}\hspace{2pt}\graphNred{.5}\hspace{2pt}\rangle_\mathrm{}. 
\end{equation}
The identity~\eqref{eq:id6} is the same as in the standard $QSSEP$ (see Ref.~\cite{bauer2019equilibrium}). 
However, we anticipate that for $\nu=2$, more complicated moments of the 
correlator will satisfy only a subset of the relationships obtained 
in~\cite{bauer2019equilibrium}. This is expected because for $\nu=2$ only red diagrams with 
only two vertices are allowed, whereas for $\nu\to\infty$ diagrams with an arbitrary number 
of vertices can be present. 

Finally, let us now consider the last diagram in~\eqref{eq:d3},  
which violates the momentum conservation at the two vertices. 
From~\eqref{eq:M6} one obtains the relations 
\begin{equation}
	\label{eq:weird2}
	\langle\hspace{2pt}\graphFdeuxr{.5}\hspace{2pt}\rangle_\mathrm{}
	\rightarrow   \hspace{4pt}\langle\hspace{2pt}
	\graphNred{.5}
	\hspace{4pt}\graphNred{.5}\hspace{3pt}\rangle_\mathrm{}
	=-\hspace{6pt} \langle\hspace{2pt}\graphFdeuxred{.5}\hspace{2pt}\rangle_\mathrm{}. 
\end{equation}
Now, at each vertex the number of $0,1$ is not preserved,  suggesting 
that at a vertex violating the momentum conservation 
the conservation of $\rho,\sigma$  is also violated. 
As we will discuss in the following, Eq.~\eqref{eq:weird2} is understood 
in the framework of random matrix theory. 

For the following, it is convenient to change the notation  
denoting with a directed numbered line the correlator $G_{kq}^{\rho\sigma}$ as
\begin{equation}
	G_{kq}^{\rho\sigma}:= \graphline{.4},\quad a=\rho+2\sigma. 
\end{equation}
This allows to rewrite the identities~\eqref{eq:id4}~\eqref{eq:id5}~\eqref{eq:id6} 
as 
\begin{align}
	\label{eq:two-1}
	\langle\graphNtwofin{.5}{0}{3}\rangle& = 
	\langle\graphNtwofin{.5}{0}{0}\rangle\\
	\label{eq:two-2}
	\langle\graphNFdeuxfin{.5}{0}{0}\rangle &= \langle
	\graphNFdeuxfin{.5}{2}{1}\rangle\\
	\label{eq:two-id}
	\langle\graphNnodefin{.5}{black}{0}{0}\rangle_\mathrm{}&=
	\langle\graphNnodefin{.5}{black}{1}{2}\rangle_\mathrm{}+
	\langle\graphNnodefin{.5}{black}{0}{3}\rangle_\mathrm{}
\end{align}
where one can exchange $0\leftrightarrow3$ and $1\leftrightarrow2$. 
The identity~\eqref{eq:weird2} becomes 
\begin{equation}
	\label{eq:w-1}
	\langle\graphFdeuxfin{.5}{3}{0}\rangle= -
	\langle\graphFdeuxfin{.5}{1}{2}\rangle
\end{equation}
To understand the behavior of higher moments of $G_{kq}$ let us now 
consider  $\prod_{j=1}^3\langle G_{k_jq_j}^{\rho_j\sigma_j}\rangle$. 
As before, to have a nonzero steady-state value, each momentum has to appear an 
even number of times. This means that the maximum number of independent 
momenta is three. Let us denote as $\mathcal{C}_3$ the set of diagrams 
with three independent momenta. We have 
\begin{equation}
	\label{eq:tre}
	\mathcal{C}_3=\Big\{\graphFktroislabel{.3}{}{}{},
	\graphFtroisllabel{.3}{}{}{},
	\graphNFdeuxllabel{.5}{}{}{},
	\graphNFdeuxlrlabel{.5}{}{}{},
\graphNtwolxx{.5}{}{}{}\Big\}.
\end{equation}
The diagrams $\mathcal{C}_2$  with only two independent momenta are  
\begin{equation}
	\label{eq:due}
	\mathcal{C}_2=\Big\{\graphNpointFdeuxllabel{.5}{}{}{},
	\graphNpointFdeuxlrlabel{.5}{}{}{},
	\graphNpointFdeuxlrrlabel{.5}{}{}{},
	\graphNnodefin{.5}{black}{}{}
\graphNparlabel{.5}{}\Big\}. 
\end{equation}
Finally, if all the momenta are equal we obtain 
\begin{equation}
	\label{eq:una}
	\mathcal{C}_1=\Big\{\graphNtroislabel{.5}{}{}{} \Big\}. 
\end{equation}
Notice that in both~\eqref{eq:tre} 
and~\eqref{eq:due} there are diagrams that do not satisfy the 
momentum conservation at all the vertices. 
To determine the allowed values for the indices $\rho_j,\sigma_j$, 
which give nonzero steady-state moments,  one 
has to write the generalization of~\eqref{eq:evol-square}. 
One obtains  a system of $20$ equations. 
For the diagrams in class $\mathcal{C}_3$ we obtain the following 
relationships 
\begin{align}
\label{eq:w-2}
&\left\langle\graphFktroislabel{.3}{0}{0}{3}\right\rangle=-
	\left\langle\graphFktroislabel{.3}{0}{2}{1}\right\rangle= 
\left\langle\graphFktroislabel{.3}{1}{0}{2}\right\rangle=-\left\langle\graphFktroislabel{.3}{1}{2}{0}
\right\rangle\\\nonumber\\\label{eq:w-2a}
	& \left\langle\graphFtroisllabel{.3}{0}{0}{0}\right\rangle=
	\left\langle\graphFtroisllabel{.3}{0}{2}{1}\right\rangle\\
	& \Big\langle\graphNFdeuxllabel{.5}{0}{0}{0}\Big\rangle
	=\Big\langle\graphNFdeuxllabel{.5}{0}{2}{1}\big\rangle= 
	\Big\langle\graphNFdeuxllabel{.5}{0}{3}{3}\Big\rangle\\
	\label{eq:w-3}
	&\Big\langle\graphNFdeuxlrlabel{.5}{0}{0}{3}\Big\rangle = -\Big\langle
	\graphNFdeuxlrlabel{.5}{0}{2}{1}\Big\rangle\\
	& \Big\langle\graphNtwolxx{.5}{0}{0}{3}\Big\rangle=\Big\langle
	\graphNtwolxx{.5}{0}{0}{0}\Big\rangle. 
\end{align}
Notice that here we do not show diagrams that are trivially related 
to the ones above. For instance, from the leftmost diagram 
in~\eqref{eq:w-2}  one obtains an equivalent one by exchanging $0\to 3$. 
In the last diagram~\eqref{eq:w-3}  one can exchange $0\leftrightarrow 3$. 
Again, the diagrams above have vertices with only two momentum lines. 
Except for the diagrams that violate momentum conservation, the identities above 
imply that for each diagram in~\eqref{eq:tre}, all the configurations of 
indices $\rho_j\sigma_j$ give the same steady-state value. 
For the diagrams with two independent momenta, we obtain 
\begin{align}
	\label{eq:firs}
	& \Big\langle\graphNpointFdeuxlrrlabel{.5}{0}{3}{0}\Big\rangle=-
	\Big\langle\graphNpointFdeuxlrrlabel{.5}{0}{1}{2}\Big\rangle\\\nonumber\\
	\label{eq:sec}
	&\Big\langle
	\graphNpointFdeuxllabel{.5}{0}{0}{0}\Big\rangle = 
	\Big\langle\graphNpointFdeuxllabel{.5}{1}{2}{0}\Big\rangle\\\nonumber\\
	\label{eq:th}
	&\Big\langle\graphNpointFdeuxllabel{.5}{0}{0}{3}\Big\rangle = 
	\Big\langle\graphNpointFdeuxllabel{.5}{2}{1}{0}\Big\rangle\\\nonumber\\
	\label{eq:fou}
	& \Big\langle\graphNpointFdeuxllabel{.5}{0}{1}{2}\Big\rangle= 
	\Big\langle\graphNpointFdeuxllabel{.5}{0}{0}{0}\Big\rangle- 
	\Big\langle\graphNpointFdeuxllabel{.5}{2}{1}{0}\Big\rangle
\end{align}
The last identity~\eqref{eq:fou} is obtained by applying the identity~\eqref{eq:id6} 
to the left vertex, whereas the other constraints are similar to 
the ones for the diagrams in $\mathcal{C}_3$. 
The only diagram left with two independent momenta gives the relationships 
\begin{align}
	\label{eq:uno}
	&\Big\langle\graphNnodelabel{.5}{0}{0}
	\graphNparlabel{.5}{0}\Big\rangle= \Big\langle\graphNnodelabel{.5}{0}{0}
	\graphNparlabel{.5}{3}\Big\rangle\\\nonumber\\
	\label{eq:due}
	&\Big\langle\graphNnodelabel{.5}{0}{3}
	\graphNparlabel{.5}{0}\Big\rangle=\Big\langle\graphNnodelabel{.5}{0}{0}
	\graphNparlabel{.5}{0}\Big\rangle-\Big\langle\graphNnodelabel{.5}{1}{2}
	\graphNparlabel{.5}{0}\Big\rangle. 
\end{align}
From the first diagram in~\eqref{eq:uno}, i.e., the one with all the labels equal, 
one can obtain an equivalent one by exchanging $3\leftrightarrow 0$. 
The same holds for  the second diagram in~\eqref{eq:uno}. Again, the identities 
in~\eqref{eq:uno} and~\eqref{eq:due} are obtained by applying~\eqref{eq:id6}. 
Finally, the diagram with all the momenta equal, i.e., in $\mathcal{C}_1$ 
(cf.~\eqref{eq:una}), satisfies the constraint 
\begin{equation}
	\label{eq:three-id}
	\Big\langle\graphNtroislabel{.5}{0}{0}{0}\Big\rangle= 
	\Big\langle\graphNtroislabel{.5}{0}{0}{3}\Big\rangle + 2 
	\Big\langle\graphNtroislabel{.5}{1}{2}{0}\Big\rangle.  
\end{equation}
Now, Eq.~\eqref{eq:three-id} is the same as in the 
standard $QSSEP$~\cite{bauer2019equilibrium}. However, in the 
original $QSSEP$ there are three nontrivial relationships 
involving the third moment of the correlator, whereas for the $\nu$-$QSSEP$ with 
$\nu=2$ only~\eqref{eq:three-id} survives. The reason is, again,  that for 
$\nu=2$ the indices $\rho_j,\sigma_j$ can take only the values $0,1$.

Before concluding, to illustrate the type of constraints that one obtains for higher moments, 
we mention that for the fourth moment $\prod_{j=1}^4G^{\rho_j\sigma_j}_{kk}$ with only one 
independent momentum $k$ there are three identities as 
\begin{align}
	\label{eq:f-1}
	& 3\Big\langle\graphNquatrlabel{.5}{1}{2}{1}{2}\Big\rangle=\Big\langle\graphNquatrlabel{.5}{0}{0}{0}{0}
	\Big\rangle-
	4\Big\langle\graphNquatrlabel{.5}{0}{0}{0}{3}\Big\rangle+	3\Big\langle
	\graphNquatrlabel{.5}{0}{0}{3}{3}\Big\rangle\\\nonumber\\
	\label{eq:f-2}
	& 6\Big\langle\graphNquatrlabel{.5}{0}{3}{1}{2}\Big\rangle=\Big\langle
	\graphNquatrlabel{.5}{0}{0}{0}{0}\Big\rangle+
	2\Big\langle\graphNquatrlabel{.5}{0}{0}{0}{3}\Big\rangle-3\Big\langle
	\graphNquatrlabel{.5}{0}{0}{3}{3}\Big\rangle\\\nonumber\\
	\label{eq:f-3}
	& 3\Big\langle\graphNquatrlabel{.5}{0}{0}{1}{2}\Big\rangle=\Big\langle
	\graphNquatrlabel{.5}{0}{0}{0}{0}\Big\rangle-
	2\Big\langle\graphNquatrlabel{.5}{0}{0}{0}{3}\Big\rangle. 
\end{align}
Eqs.~\eqref{eq:f-1}\eqref{eq:f-2}\eqref{eq:f-3} are obtained by deriving 
the generalization of~\eqref{eq:gener} and solving for the zero eigenvectors. 
An explicit calculation shows that one has to solve a system of $924$ equations. 

The identities derived above are a subset of the constraints 
that hold for the fourth moment of the real-space correlator in 
the $QSSEP$~\cite{bauer2019equilibrium}. 
Precisely, by using the results of Ref.~\cite{bauer2019equilibrium}, one can obtain a 
set of $11$ identities for the fourth moment 
$\langle\prod_{j=1}^4 G^{\rho_j\sigma_j}\rangle$. 
The corresponding diagrams contain  more than two vertices. 
However, one can show that the subset of diagrams with 
only two vertices, satisfy~\eqref{eq:f-1}~\eqref{eq:f-2}~\eqref{eq:f-3}. 

%############################################
\subsection{Role of the conserved quantities}
\label{sec:cons}

A crucial question is whether it is possible to determine all the nonzero steady-state moments 
of the correlators from initial-state data. For the two-point correlator this is the case because 
it can be written in terms of the initial occupations of the momentum modes $N_k$  
via~\eqref{eq:Nk}. This holds true for any $\nu$, as stressed in Section~\ref{sec:f-corr}. 
Let us now discuss how to determine the steady-state expectation values of 
the second moments $\langle G_{k_1q_1}^{\rho_1\sigma_1}G_{k_2q_2}^{\rho_2\sigma_2}\rangle$. 
By using~\eqref{eq:Gk-evol}, one can build three conserved quantities 
$N_1^{kq},N_2^{kq},N_3^{kq}$ defined as  
\begin{align}
	\label{eq:c}
	& N_1^{kq}=\mathrm{Tr}(G^{\rho\sigma}_{kk})\mathrm{Tr}(G^{\rho\sigma}_{qq}), \\
	\label{eq:c1}
	& N_2^{kq}=\mathrm{Tr}\Big
	(\sum_{\sigma_1}G^{\rho_1\sigma_1}_{kq}G^{\sigma_1\rho_2}_{qk}\Big),\\
	\label{eq:c2}
	& N_3^{kq}=G_{kq}^{\rho\rho}G_{kq}^{\sigma\sigma}
	-G_{kq}^{\rho\sigma}G_{kq}^{\sigma\rho}, 
\end{align}
where the traces is over the indices of the unit cell. These quantities are preserved even at 
the level of the single realization of the noise  
$dW_t^{\rho}$, as one can verify by using~\eqref{eq:Gk-evol}. 
Notice that the last term in~\eqref{eq:c} is the determinant of $G_{kq}^{\rho\sigma}$. 
Now, it is straightforward to rewrite $N_j^{kq}$ as  
\begin{align}
	\label{eq:c2-1}
	& N^{kq}_1=
	2\langle\graphNtwofin{.5}{0}{3}\rangle_\mathrm{}+
	2\langle\graphNtwofin{.5}{0}{0}\rangle_\mathrm{}\\
	\label{eq:c2-2}
	& N^{kk}_1=2\langle\graphNnodefin{.5}{black}{0}{3}\rangle_\mathrm{}+
	2\langle\graphNnodefin{.5}{black}{0}{0}\rangle_\mathrm{}\\
	\label{eq:c2-3}
	& N^{kq}_2=2\langle\graphNFdeuxfin{.5}{0}{0}\rangle_\mathrm{} +2\langle
	\graphNFdeuxfin{.5}{2}{1}\rangle_\mathrm{}\\
	\label{eq:c2-4}
	& N^{kk}_2=
	2\langle\graphNnodefin{.5}{black}{0}{0}\rangle_\mathrm{}+
	2\langle\graphNnodefin{.5}{black}{1}{2}\rangle_\mathrm{}\\
	\label{eq:c2-5}
	& N^{kq}_3=
	\langle\graphFdeuxfin{.5}{3}{0}\rangle_\mathrm{}-
	\langle\graphFdeuxfin{.5}{2}{1}\rangle_\mathrm{},
\end{align}
Finally, by inverting~\eqref{eq:c2-1}-\eqref{eq:c2-5} 
one can extract the steady state values of the correlators 
in terms of $N_j^{kq}$. Similar relationships were derived for the 
standard $QSSEP$~\cite{bauer2019equilibrium}. Now, it is natural to conjecture 
that, even for generic $\nu$,  the same strategy can be employed to obtain 
the steady-state expectation value of arbitrary moments of the correlators 
in terms of initial-state expectation values.

%############################################
\subsection{Statistics of steady-state correlators} 
\label{sec:symm}

As for the  $QSSEP$~\cite{bauer2019equilibrium}, the identities derived in the 
previous sections reflect the Haar invariance  of the statistics of 
steady-state correlators. To highlight that, following Ref.~\cite{bauer2019equilibrium}, 
we consider the functions $Z_n(A)$ as 
\begin{equation} 
	\label{eq:Z}
	Z_n(A)=\langle (\mathrm{Tr}\,GA)^n\rangle,
\end{equation}
where the trace is over the both unit cell indices and momentum indices. $Z_n$ are 
the coefficients of the generating function $Z(A)=\exp({\mathrm{Tr}GA})$. 
By taking appropriate derivatives with respect to the matrix elements 
$A_{kq}^{\rho\sigma}$ one obtains all the moments of $G_{kq}^{\rho\sigma}$. 
For $n=1$, one has  
\begin{equation} \left\langle \sum_{k
q,\rho\sigma}G_{kq}^{\rho\sigma}A_{qk}^{\sigma\rho}\right\rangle=\frac{1}{2}
\sum_{k,\rho}\mathrm{Tr}(\langle G_{kk})\rangle A_{kk}^{\rho\rho} 
\end{equation}
For $n=2$, we can expand $\langle [\mathrm{Tr}GA]^2\rangle$ in terms of the 
diagrams in~\eqref{eq:d3}. 
All the possible diagrams appear in the expansion. 
However, by using the identities~\eqref{eq:two-1}~\eqref{eq:two-2} 
and~\eqref{eq:two-id}, one obtains that 
\begin{multline} \label{eq:sym-3}
	\langle[\mathrm{Tr}(GA)]^2\rangle= 
	\sum_{k\ne q} \Big[
		\langle\graphNFdeuxfin{.5}{1}{2}\rangle 
		\mathrm{Tr}(A_{qk} A_{kq})+
		\langle\graphFdeuxfin{.5}{0}{3}\rangle(\mathrm{Tr}
 A_{qk})^2 \\+
\langle \graphFdeuxfin{.5}{1}{2} \rangle
		\mathrm{Tr}(A_{qk} A_{qk})+
		\langle\graphNtwofin{.5}{0}{3}\rangle
		\mathrm{Tr}( A_{qq})\mathrm{Tr}( A_{kk})\Big]+\\
		\sum_{k}\Big[
			\langle
			\graphNnodefin{.5}{black}{1}{2}\rangle \mathrm{Tr}(A_{kk} 
			A_{kk})+\langle\graphNnodefin{.5}{black}{0}{3}\rangle
	(\mathrm{Tr}A_{kk})^2\Big], 
\end{multline}
where the trace is over the indices $\rho,\sigma$.  
All the prefactors of the traces in~\eqref{eq:sym-3} can be 
rewritten in terms of the conserved quantities 
in~\eqref{eq:c2-1}-\eqref{eq:c2-5}. 

Now, the crucial observation is that the terms that survive the 
noise average in Eq.~\eqref{eq:sym-3} are the ones that are 
invariant under the transformation 
\begin{equation} 
	\label{eq:su2-t}
	A_{kq}\to V_{k}^\dagger A_{kq}V_q 
\end{equation}
with $V_k,V_q$ random $SU(2)$ matrices. Notice that the terms 
$[\mathrm{Tr}(A_{qk})]^2$ and $\sum_{\rho\sigma}(A^{\rho\sigma}_{qk} 
A^{\sigma\rho}_{qk})$ are not 
separately invariant under~\eqref{eq:su2-t}. However, the combination appearing 
in~\eqref{eq:sym-3} is invariant due to the identity~\eqref{eq:w-1}, and it is 
the determinant of $A_{kq}$. One can check that a similar result holds 
for larger $n$ and generic $\nu$. It is in general challenging to construct analytically $Z(A)$ 
for generic $SU(\nu)$. Moreover, 
for small $\nu$ a direct computation of the noise average is convenient, 
as we show in the following. 

The invariance under~\eqref{eq:su2-t} implies that the 
steady-state noise average of a generic moment  
$\prod_{j=1}^n G_{k_jq_j}^{\rho_j\sigma_j}$ with fixed $k_j,q_j,\rho_j,\sigma_j$ 
can be computed as 
\begin{equation}
	\label{eq:trace}
	\Big\langle\prod_{j=1}^n G_{k_jq_j}^{\rho_j\sigma_j}\Big\rangle 
	=\int\!\! d\eta(V)
	\prod_{l}[V^\dagger_{k_l}G_{k_lq_l}V_{q_l}]^{\rho_l\sigma_l}, 
\end{equation}
where $V_k$ are random $SU(2)$ matrices, and we denote 
by $d\eta(V)$ the corresponding Haar measure. 
Precisely, a random  $SU(2)$ matrix $V_k$ can be parametrized as 
\begin{equation}
	V_k=\left(\begin{array}{cc}
			\rho_k e^{i\phi_k} & e^{-i\psi_k}\sqrt{1-\rho_k^2}\\
			-e^{i\psi_k}\sqrt{1-\rho_k^2} & e^{-i\phi_k}\rho_k
		\end{array}
	\right)
\end{equation}
with $\phi_k,\psi_k\in[0,2\pi]$, and $0\le\rho_k\le1$. 
The average with the  Haar measure over $V_k$ is written as 
\begin{equation}
	\int d\eta(V_k)=\int_{0}^1 2\rho_kd\rho_k \int_0^{2\pi}\frac{d\phi_k}{2\pi}\int_0^{2\pi}
	\frac{d\psi_k}{2\pi}. 
\end{equation}
The integral in~\eqref{eq:trace} can 
be written in terms of quantities that are invariant under the 
Haar average. Similar decomposition rules exist for 
generic $SU(N)$~\cite{creutz1978on,weingarten1980non}. 
Precisely, we employ the key formula 
\begin{multline}
	\label{eq:form1}
	\int \!\!d\eta (V) V_{i_1j_1}\cdots V_{i_{2k}j_{2k}}=\\
\frac{1}{(k+1)!}\sum_{\mathrm{pairings}\,P} \prod_{(a,b)\in P}\varepsilon_{i_a,i_b}
\varepsilon_{j_a,j_b},
\end{multline}
where $\varepsilon_{ij}$ is the completely antisymmetric two-dimensional 
tensor  with $\varepsilon_{12}=1$, and the sum is over all the possible ways of 
pairing the indices. 
Clearly, the Haar average of the product of an odd number of terms $V_{ij}$ vanishes. The 
average of the product of an even number of terms $V^*_{ij}$ is obtained from~\eqref{eq:form1} 
by using that $V^{*}_{ij}=\varepsilon_{ip}V_{ps}\varepsilon^{-1}_{sj}$, where repeated 
indices are summed over, and $\varepsilon^{-1}_{sj}=\varepsilon_{js}$. 
Let us consider the general case of a product of $n+m$ $SU(2)$ random matrices 
containing both $V$ and $V^*$ terms. One has 
\begin{multline}
	\label{eq:form2}
	\int\!\! d\eta (V) V_{i_1j_1}\cdots V_{i_nj_n} 
	V^*_{i_{n+1}j_{n+1}}\cdots V^*_{i_{n+m}j_{n+m}}=\\
	\frac{1}{[(n+m)/2+1]!}\sum_{\mathrm{pairings}\,P} \prod_{(a,b)\in P}K_{i_a,j_a,i_b,j_b}, 
\end{multline}
where $K_{i_a,j_a,i_b,j_b}=\varepsilon_{i_ai_b}\varepsilon_{j_aj_b}$ for $VV$ and $V^*V^*$ pairings, 
and $K_{i_a,j_a,i_b,j_b}=\delta_{i_ai_b}\delta_{j_aj_b}$ for $VV^*$ pairings. 
Notice that for the Haar average~\eqref{eq:form2} to be nonzero $n+m$ has to be 
even. One can now apply~\eqref{eq:form1} and~\eqref{eq:form2} to the generic 
steady-state value of $\langle G_{k_1q_1}G_{k_2q_2}\cdots G_{k_nq_n}\rangle$. 
Indeed, the result is obtained  by performing the Haar average of 
$[V_{k_1}^\dagger G_{k_1q_1}V_{q_1}]^{\rho_1\sigma_1}
[V^\dagger_{k_2}G_{k_2q_2}V_{q_2}]^{\rho_2\sigma_2}\cdots 
[V^\dagger_{k_n}G_{k_n q_n}V_{q_n}]^{\rho_n\sigma_n}$ over the $SU(2)$ random matrices 
$V_{k_j},V_{q_j}$. This can be easily done by first grouping the matrices $V_k$ 
with the same momentum label and then using~\eqref{eq:form1} 
and~\eqref{eq:form2}, exploiting the fact that the Haar averages 
over random matrices with different momentum labels factorize. 
Notice that for the pairings involving $V^\dagger_{k_j}$ 
one has to take a matrix transposition in~\eqref{eq:form1} and~\eqref{eq:form2}. 
It is natural to expect that the results above generalize to arbitrary $\nu$, although 
the identities~\eqref{eq:form1} and~\eqref{eq:form2} become more cumbersome. 

Let us now discuss the consequences of Haar invariance on steady-state expectation 
values of arbitrary functions of the real-space correlators. 
Precisely, let us consider a generic function $\mathcal{F}(z)$ of 
$G_{ij}$. This includes functions of the correlator $G_A$ restricted to a subregion $A$, such 
as the entanglement entropy. It is clear that the steady-state value of 
$\mathrm{Tr}{\mathcal F}(G)$ is obtained as 
\begin{equation}
	\langle\mathrm{Tr}{\mathcal F}(G)\rangle=
	\int_{\mathscr{E}} d\eta(V) \mathrm{Tr}{\mathcal F}(V^\dagger G^{(0)}V),
\end{equation}
where $G^{(0)}$ is the initial correlator, and $\mathscr{E}$ is the ensemble of $L\times L$ 
random matrices obtained as the inverse Fourier transform of random block diagonal 
$SU(\nu)$ matrices $V_{k}$ as 
\begin{equation}
	\label{eq:random}
	V_{k}=\mathds{1}_{L/\nu} \otimes M_k, 
\end{equation}
where $M_k$ is a random $SU(\nu)$ matrix. For $\nu=1$, $V_{k}$ is diagonal matrix 
with a random momentum-dependent phase on the diagonal. A similar result holds for 
the $QSSEP$~\cite{bauer2019equilibrium}, and it was used in Ref.~\cite{bernard2021entanglement} to 
determine the steady-state statistics of the entanglement.

%############################################
\section{Full-time entanglement dynamics for $\nu=1$} 
\label{sec:e-dyn}

In the following we show that for $\nu=1$ it is possible to 
obtain the full-time dynamics or arbitrary entanglement-related quantities in the 
space-time scaling limit $t,\ell\to\infty$ with $t/\ell^2$ fixed. 
Although here we restrict ourselves to $\nu=1$, it should be possible to generalize our 
results to $\nu>1$, at least for moderate $\nu$. 
The starting point are the results of Section~\ref{sec:nu-1}. 
Let us start by computing the moments $M_n$ of the reduced 
correlation matrix defined as 
\begin{equation}
	\label{eq:moments}
	M_n= \mathrm{Tr}(\langle G_A^n\rangle). 
\end{equation}
To illustrate the strategy of the derivation, let us consider the 
case with $n=2$. One has to determine the large $\ell,t$ behavior of the 
expression  
\begin{multline} 
	\label{eq:2-mom}
	M_2=\\
	\frac{1}{L^{2}}\sum_{n,m}\sum_{\substack{k_1,q_1\\
	k_2,q_2}}e^{-i(k_1-q_2)n+i(q_1-k_2)m}
	\langle G_{k_1q_1}G_{k_2q_2}\rangle,  
\end{multline}
where $\langle G_{k_1q_1}G_{k_2q_2}\rangle$ evolves with time, except for 
$k_1=q_1$ and $k_2=q_2$, or  $k_1=q_2,k_2=q_1$, which correspond to conserved 
quantities. In the following we will consider the thermodynamic limit $L\to\infty$, 
replacing $1/L\sum_k\to\int_{-\pi}^\pi dk/(2\pi)$.  

To perform the sum over $n,m\in[1,\ell]$ we use the trivial identity 
as~\cite{calabrese2012quantum}  
\begin{equation} 
\label{eq:id} 
\sum_{m=1}^\ell e^{ i m k}=
\frac{\ell}{4}\int_{-1}^1d\xi w(k) e^{i(\ell\xi+\ell+1)k/2},
\end{equation}
with 
\begin{equation}
w(k)=\frac{k}{\sin(k/2)}. 
\end{equation}
Here we focus on the quench dynamics from initial states that exhibit two-site 
translation invariance, such as the N\'eel and the dimer state (see~\eqref{eq:neel} and~\eqref{eq:dimer}). 
This means that $G_{kq}$ is almost diagonal. Precisely, here we focus on 
$G_{kq}$ of the form 
\begin{equation} 
	\label{eq:dimer-ft}
	G_{kq}= \frac{1}{2}f(k)\delta_{kq} + \frac{1}{2}g(k)\delta_{k,q+\pi}, 
\end{equation}
with $g(k-\pi)=g^*(k)$, to ensure Hermiticity. 
The case of the N\'eel state corresponds to $f(k)=g(k)=1$, whereas 
the dimer state corresponds to $f(k)=1+\cos(k)$ and $g(k)=-i\sin(k)$. 
The evolution of the correlator $\langle G_{k_1q_1}G_{k_2q_2}\rangle$ is given 
as (cf.~\eqref{eq:four-nu-1}) 
\begin{equation} 
\label{eq:inte}
\frac{d}{dt}\langle
G_{k_1q_1}G_{k_2q_2}\rangle= \varepsilon_{k_1q_1k_2q_2} \langle
G_{k_1q_1}G_{k_2q_2}\rangle. 
\end{equation}
The ``energies'' $\varepsilon_{k_1q_1k_2q_2}$ can take only few values because 
$G_{kq}$ is almost diagonal. Precisely, one has  
\begin{equation} 
	\label{eq:vel-1}
	\varepsilon_{k_1q_1k_2q_2}=-8\gamma(1+\cos(k_1-k_2)), 
\end{equation}
for $q_1=k_1\pm \pi$ and $q_2=k_2\pm\pi$, and  
\begin{equation}
	\label{eq:vel-2}
\varepsilon_{k_1q_1k_2q_2}=-4\gamma  
\end{equation}
for $q_1=k_1\,\,\mathrm{or}\,\,q_2=k_2$. 
The solution of~\eqref{eq:vel-1} is straightforward and 
it gives an exponential decay, which can be used in~\eqref{eq:2-mom}. 
We report the detailed calculation  for $n=2$ in Appendix~\ref{app:1}. 
Let us discuss the general strategy. For generic $n$, $M_n$ is determined as 
\begin{multline} \label{eq:gen-form} 
	M_n=
	\left(\frac{\ell}{4}\right)^n\prod_{j=1}^n
	\int\frac{dk_j}{2\pi} \frac{dq_j}{2\pi} 
	\\\times \prod_{j'=1}^n\int d\xi_{j'}
	e^{i\ell\sum_{p=1}^n\xi_p(q_{p-1}-k_p)/2}
	\\\times \prod_{p=1}^n
	w(q_{p-1}-k_p)G_{k_p,q_p} e^{t\varepsilon(\{k_j,q_j\})}, 
\end{multline}
where $w(x)$ is defined in~\eqref{eq:id}, and the energies 
are given by~\eqref{eq:vel-1} and~\eqref{eq:vel-2} for $n=2$. 
For $n>2$, $\varepsilon_{k_jq_j}$ can be determined by considering 
$d\langle G_{k_1q_1}G_{k_2q_2}\cdots G_{k_nq_n}\rangle$ and 
applying~\eqref{eq:dyn} to each of the $G_{k_jq_j}$ 
and~\eqref{eq:four-nu-1} to all the pairs $G_{k_rq_r}G_{k_sq_s}$. 
In writing the phase in~\eqref{eq:gen-form} we used the fact that 
for generic one-site and two-site translation invariant initial states 
one has $G_{k_p,q_p}\propto\delta_{k_p-q_p,0\,\,\mathrm{mod}\,\pi}$, which 
allows us to neglect the term $\ell+2$ in the complex 
exponentials~\eqref{eq:id} because it would give a trivial phase. 
Let us now consider the large $\ell$ limit. One can check that 
the integration over $\xi_j$ gives a nonzero contribution only 
if there is an even number of momenta $q_j$ such that 
$q_j=k_j\pm \pi$ and the total $\pi$ shift of the $q_j$ vanishes. 
From now on we define two momenta $k_j,k_{j'}$ paired if $q_j=k_j\pm \pi$ 
and $q_j'=k_{j'}\mp \pi$. 
The reason why one can restrict  to the terms with an even 
number of paired momenta is that the integration over $\xi_j$ gives
in the large $\ell$ limit a factor $\delta(q_{j-1}-k_j)$. 
One can check that the ``chain'' of delta functions 
arising from the integrals over all the $\xi_j$ 
gives a nonzero result only in the presence of an 
even number of paired  momenta. 
It is important to stress that terms that do not have an even number 
of paired momenta give a subleading contribution in the limit $\ell,t\to\infty$, 
although they are non zero, except for the quench from the N\'eel state, 
for which they vanish. We anticipate (see Fig.~\ref{fig:trneel}) 
that this is the reason why for the N\'eel state 
finite-size corrections are smaller.

Let us now discuss the structure of the energies $\varepsilon$. It is
easy to check that they depend in a simple way on the paired momenta. 
For the case with $2p$ paired momenta one has 
\begin{equation}
	\label{eq:quasi-e}
	\varepsilon=-8\gamma(p+\sum_{\mathrm{paired}\,j,j'}\cos(k_j-k_{j'})),
\end{equation}
where the sum is over all the paired $k_j,k_{j'}$ momenta. 
In the following we consider the case in which at least two momenta 
are paired, i.e., we neglect the momenta configurations with $k_j=q_j$ for any 
$j$, which do not evolve with time. 
Now, the fact that momenta have to be paired implies that the 
phase factor $e^{i\ell\sum_{p}\xi_p(q_{p-1}-k_p)}$  contain $\pi$ shifts. 
It is convenient to remove them by changing variables as follows. For the 
case with $k_j$ and $k_{j'}$ paired one can remove the $\pi$ shifts 
by redefining $k'_{l}=k_{l}\mp\pi$ for $l\in[j+1,j']$.
The same change of variables can be iterated to 
remove the $\pi$ shifts associated with all the paired momenta. This 
change of variables will lead to some minus signs in the 
energies $\varepsilon$ and in the $G_{kq}$. 

It is also convenient to change variables in~\eqref{eq:gen-form} defining 
$\zeta_j$ as 
\begin{equation}
	\label{eq:zeta-def}
	\zeta_0=\xi_1,\quad \zeta_j=\xi_{i+1}-\xi_i,\quad i\in[1,n-1]. 
\end{equation}
Now, we can rewrite~\eqref{eq:gen-form} as  
\begin{multline} \label{eq:gen-form-1} M_n\simeq 
	\left(\frac{\ell}{4}\right)^n\prod_{j=1}^n\int\frac{dk_j}{2\pi}
	\prod_{j'=1}^{n-1}\\\int d\zeta_{j'} \mu(\{\zeta_j\})
	e^{i\ell\sum_{p=1}^{n-1}\zeta_p(k_{p}-k_{n})/2}\prod_{p=1}^n
	w(k_{p-1}-k_p)\\
\sum_\mathrm{pairings}\left[G'_{k_1,q_1}G'_{k_2,q_2}\cdots G'_{k_n,q_n}
e^{t\varepsilon'(\{k_j,q_j\})}\right], 
\end{multline}
where we used the fact that the integrand does not depend on $\zeta_0$, and in the 
second row we sum over all the possible ways of paring the momenta $k_j$. In~\eqref{eq:gen-form-1} the  
$\simeq$ is to stress that we are ignoring the terms that do not contain paired 
momenta, and hence do not evolve with time. We will restore 
these contributions at the end of the calculation. 
The prime in $G'_{k_jq_j}$ and in $\varepsilon'$ is to take into account 
the effect of redefining $k_j$ to remove the $\pi$ shifts. From~\eqref{eq:dimer-ft} 
one has that each $G_{kq}$ contributes with a factor $f(k)$ for each momentum that 
is not paired and a factor $g(k)$ for each 
paired one. Moreover, at least for the N\'eel state and the dimer state the 
change of variable to remove the $\pi$ shifts does not have any effect on $f(k)$, whereas 
it leads to a change of sign in $g(k)$. 
In~\eqref{eq:gen-form-1} $\mu(\{\zeta_j\})$ is the result of the 
integration over $\zeta_0$, and is defined as 
\begin{multline}
\label{eq:mu}
\mu(\{\zeta_j\})=\max\Big[0,\min_{j\in[0,n]}\Big(1-\sum_{l=1}^j\zeta_l\Big)\\
+\min_{j\in[0,n]}\Big(1+\sum_{l=1}^j\zeta_l\Big)\Big], 
\end{multline}
Now, in the large $\ell$ limit one can use the stationary phase approximation~\cite{wong2001asymptotic}. 
The stationarity condition with respect to $\zeta_j$ gives that $k_j\to
k_n$ for any $j$. In this limit we can replace $w(k_{p-1}-k_p)\to 2$. Moreover,
we can also replace $k_j\to k_n$ in the product $G'_{k_1,q_1}G'_{k_2,q_2}\cdots
G'_{k_n,q_n}$. However, we have to keep distinct the momenta
in the exponents in~\eqref{eq:gen-form-1}, although  we can expand
$\cos(k_j-k_l)\approx 1-(k_j-k_l)^2/2$. 

Let us discuss how to proceed focussing on the contributions with
two paired momenta.  For instance, let us consider the case with 
$k_j$ and $k_{j'}$ ($j<j'$) paired as 
\begin{equation} q_j=k_j\pm\pi,\quad q_{j'}=k_{j'}\mp\pi,\quad q_r=k_r\quad
r\ne j,j' 
\end{equation}
Now, Eq.~\eqref{eq:gen-form-1} becomes 
\begin{multline} \label{eq:gen-form-int-2}
	M^{(2\mathrm{p})}_n=\left(\frac{\ell}{2}\right)^n\prod_{j=1}^{n}\left(\int\frac{dk_j}{2\pi}\int
	d\zeta_{j}\right) \\
	\mu(\{\zeta_j\}) e^{i\ell\sum_{p=1}^{n-1}\zeta_p(k_{p}-k_{n})/2} 
\frac{|g(k_n)|^2}{2^n}\\
\sum_{r< r'} f^{n-2-r'+r}(k_n)f^{r'-r}(k_n+\pi)e^{-4\gamma t(k_r-k_{r'})^2}, 
\end{multline}
where the functions $f(x),g(x)$ are defined in~\eqref{eq:dimer-ft}, and the 
superscript $2\mathrm{p}$ is to stress that we restrict to the situation with 
only two paired momenta. 
Here one obtains the term $|g(k_n)|^2$ because of the condition $q_j=k_{j}\pm \pi$ 
and $q_{j'}=k_{j'}\mp \pi$ and after the change of variable on $k_j$ to remove 
the $\pi$ shifts. 
Now each integration over a momentum $k_j$ that is not paired gives 
$2/\ell\delta(\zeta_j)$, and we obtain 
\begin{multline} \label{eq:gen-form-int-2}
	M^{(2\mathrm{p})}_n=\frac{1}{2^n}\frac{\ell^3}{8}
	\sum_{j<j'}\int\frac{dk_j}{2\pi}\int
	\frac{dk_{j'}}{2\pi}
	\Big(\prod_{s=1}^{n-1}\int d\zeta_{s}\Big) \mu(\{\zeta_s\})
	\\\prod_{r\ne j,j'}\delta(\zeta_r)
	f^{n-2-j'+j}(k_n)f^{j'-j}(k_n+\pi)|g(k_n)|^2\\
	e^{-i\ell\sum_{p=1}^{n-1}\zeta_p k_{n}/2+i\ell \zeta_j
	k_j/2+i\ell\zeta_{j'}k_{j'}/2} e^{-4\gamma t(k_j-k_{j'})^2},
\end{multline}
In the last row in~\eqref{eq:gen-form-int-2} we expanded 
the exponent since $k_j\to k_n$ for any $j$ in the limit $\ell,t\to\infty$. 
The integration over the Dirac deltas is trivial.  
Furthermore, one can integrate over $k_j$ and $k_{j'}$, to obtain  
\begin{multline} 
	\label{eq:gen-form-int-4} M^{(2\mathrm{p})}_n=2^{-n}\frac{\ell^2 }{4}
	\sum_{j<j'} \int\frac{k_n}{2\pi}
	\int d\zeta_{j}\int
		d\zeta_{j'}\delta(\zeta_j+\zeta_{j'})\\
		f^{n-2-j'+j}(k_n)f^{j'-j}(k_n+\pi)\\
		 |g(k_n)|^2
		\mu(\zeta_j,\zeta_{j'}) e^{-i\ell (\zeta_j+\zeta_{j'})
		k_n/2}\frac{e^{-\ell^2\zeta_j^2/(64\gamma t)}}{4\sqrt{\pi\gamma
		t}}. 
\end{multline}
After integrating over  $\zeta_{j'}$, we obtain 
\begin{multline} \label{eq:gen-form-int-5} 
	M^{(2\mathrm{p})}_n=\frac{1}{2^n}\frac{\ell^2}{4}
	\int\frac{dk_n}{2\pi}
	\sum_{j<j'} f^{n-2-j'+j}(k_n)f^{j'-j}(k_n+\pi)
			\\|g(k_n)|^2\int\!\! d\zeta_{j}
	\mu(\zeta_{j'}=-\zeta_j)
\frac{e^{-\ell^2\zeta_j^2/(64\gamma t)}}{4\sqrt{\pi\gamma t}}, 
\end{multline}
Finally, since $\mu(\zeta_j,\zeta_{j'})=(2+\zeta_j)\Theta(-\zeta_j)+(2-\zeta_j)\Theta(\zeta_j)$ 
and $-2\le \zeta_j\le 2$, the integration over $\zeta_j$ in~\eqref{eq:gen-form-int-5} gives 
\begin{multline}
	\label{eq:In-final}
	M_{n}^{(2\mathrm{p})}=
	\frac{1}{2^{n}}\sum_{j<j'}\int\!\!\frac{dk}{2\pi}
	f^{n-2-j'+j}(k)f^{j'-j}(k+\pi)|g(k)|^2\\\times 
	\left[
	4\sqrt{\frac{\gamma t}{\pi}}
	\left(e^{-\ell^2/(16\gamma
	t)}-1\right)+\ell\, \mathrm{Erf}\left(\frac{\ell}{4\sqrt{\gamma
t}}\right)\right], 
\end{multline}
where we replaced $k_n\to k$, and $\mathrm{Erf}(x)$ is the error function. 
Notice that for $n=2$, Eq.~\eqref{eq:In-final} is the full result because 
there are only terms with two pairings, i.e., $j=1,j'=2$. 
The term in the last row in~\eqref{eq:In-final}, which encodes the dynamics 
does not depend on $k$.  This means that the quasiparticles exhibit the same 
diffusive dynamics irrespective of $k$. In contrast, 
in the standard quasiparticle picture entanglement spreads ballistically, 
and quasiparticles with different $k$ have different group 
velocities~\cite{alba2021generalized}.  

Let us now discuss the case in which more than two paired momenta in~\eqref{eq:gen-form-1}. 
First, let us focus on the contribution of the product $\prod_{j=1}^n G'_{k_jq_j}$. 
As we stressed for the case of two paired momenta, one can replace in the limit 
$\ell,t\to\infty$ $k_j\to k_n$. The contribution of the product is a polynomial 
in powers of $f(k_n),f(k_n+\pi),g(k_n),g(k_n+\pi)$. 
Actually, one can check that the contribution of the term with $2r$ pairings  is the 
prefactor of $x^{2r}$ in the  expansion around $x=0$ of $\mathrm{Tr}\, Q^{2r}(x)$, 
with the matrix $Q(x)$ defined as 
\begin{equation}
	Q(x)=\frac{1}{2}\left(\begin{array}{cc}
			f(k+\pi) & x g(k+\pi)\\
			x g(k) & f(k)
	\end{array}\right). 
\end{equation}
It is convenient to diagonalize $Q$, obtaining the eigenvalues as
\begin{multline}
	\lambda_\pm=\frac{1}{4}\Big[f(k)+f(k+\pi)\pm 
		\Big(f^2(k)+f^2(k+\pi)\\-2f(k)f(k+\pi)+
	4x^2g(k)g(k+\pi)\Big)^\frac{1}{2}\Big]. 
\end{multline}
To perform the integration over the momenta $k_j$ in~\eqref{eq:gen-form-1} 
we use the trivial formula  
\begin{multline} \label{eq:gauss} \int\!\frac{dk_j}{2\pi} e^{i\ell
	\sum_{p=1}^{n-1} \zeta_p/2(k_p-k_n)-4\gamma t(\sum_p \sigma_p
k_p)^2}= \frac{1}{4\sqrt{\pi\gamma t}}\\
e^{-\ell^2\zeta_j^2/(4\gamma t)-i\ell
k_n(\sigma_j \zeta_j+\sum_{p}\zeta_p)+i\ell/2\sum_{p\ne
j}k_p(\zeta_p+\sigma_p\zeta_j)} 
\end{multline}
In the exponential in the left-hand side 
in~\eqref{eq:gauss}, $\sigma_p=\pm1$ takes into account the signs of the 
momenta $k_j$ obtained when expanding $\varepsilon'$ (cf.~\eqref{eq:gen-form-1}) 
in the limit $k_j\to k_n$. As it is clear from 
Eq.~\eqref{eq:gauss}, a single integration over any of the paired momenta is sufficient 
to decoupled them, leaving in the exponent only linear terms in $k_j$. 
One can easily integrate over all the remaining 
$k_p\ne k_n$ in~\eqref{eq:gauss}, obtaining  
\begin{equation} 
	\label{eq:gauss-1}
	\left(\frac{2}{\ell}\right)^{n-2}\prod_{p\ne
	j}\delta(\zeta_p-\sigma_p\zeta_j). 
\end{equation}
The remaining integration over $\zeta_j$ can be easily performed. 
One obtains that all the terms with any nonzero number of paired momenta give rise to 
the same dynamics, i.e., the second row in~\eqref{eq:In-final}. Notice that although this 
might look surprising, it also happens for the quench in the deterministic chain, 
where all the terms with paired momenta 
give $\min(|v_k|t,\ell)$, which describe the ballistic propagation of entangled quasiparticles~\cite{alba2023logarithmic}. 
Unlike the case of the deterministic chain, the propagation of quasiparticles is now 
diffusive. In conclusion, based on the derivation 
for $n\le 4$ we conjecture  the general formula for $M_n$ as
\begin{multline}
	\label{eq:neel-Mn}
	M_n=\frac{\ell}{2}\int\!\! \frac{dk}{2\pi}
	\Big[\lambda_-^n+\lambda_+^n-
	2^{-n}f^n(k)-2^{-n}f^n(k+\pi)\Big]\\
	\times\left[4\sqrt{\frac{\gamma t}{\pi\ell^2}}
	\left(e^{-\ell^2/(16\gamma
t)}-1\right)+\mathrm{Erf}\left(\frac{\ell}{4\sqrt{\gamma
t}}\right)\right]\\
+\frac{\ell}{2}\int\!\! \frac{dk}{2\pi} [2^{-n}f^n(k)+2^{-n}f^n(k+\pi)].  
\end{multline}
The first term in~\eqref{eq:neel-Mn} is $\mathrm{Tr}\, Q^n(1)=\lambda_+^n+\lambda_-^n$, 
where we subtracted the terms with no paired momenta because they do not evolve, and we 
used that the terms with paired momenta exhibit the same dynamics. 

The last term in~\eqref{eq:neel-Mn} 
is the steady-state value of $M_n$. This corresponds to the momenta configurations for 
which $\varepsilon$ vanishes. For instance, for $n=2$ one has $G_{kk}G_{qq}$ and $G_{kq}G_{qk}$, 
where $k,q$ can be equal. To extract the steady-state value of $M_n$ one has to 
compute a multidimensional integral similar to~\eqref{eq:gen-form}, although without 
time-dependent terms. The large $\ell$ analysis is performed in a similar way. One obtains 
that at the leading order in the limit $\ell\to\infty$ only the terms with $k_j=q_j$ 
contribute. For instance, this means that for $n=2$ the terms with $G_{kq}G_{qk}$ do not 
contribute. This also means that the steady-state value of $M_n$, and hence the entanglement entropy, 
is obtained from $G_{kk}$, i.e., the density of fermionic modes, which defines the $GGE$ describing 
the quench in the deterministic chain. Crucially, this holds for the infinite chain, i.e, if one takes the thermodynamic 
limit $L\to\infty$ first, and then the limit $\ell\to\infty$. For instance, 
if one takes the limit $L\to\infty$ with $\ell/L$ finite, the terms originating from $G_{kq}G_{qk}$ contribute to the 
steady-state correlators. 
For the N\'eel quench we have $f(k)=f(k+\pi)=1$ and $g(k)=g(k+\pi)=1$, 
implying that $\lambda_+=1$ and $\lambda_-=0$. For the quench from the dimer 
state, we have $f(k)=1+\cos(k)$ and $g(k)=-i\sin(k)$. Again, one has 
$\lambda_+=1$ and $\lambda_-=0$. 

Eq.~\eqref{eq:neel-Mn} allows one to determine the dynamics of 
$\mathrm{Tr}{\mathcal F}(G_A)$ for a generic function $\mathcal{F}(x)$ 
that admits a Taylor expansion around $x=0$. One obtains 
\begin{multline}
\label{eq:trfga}
\mathrm{Tr}{\mathcal F}(G_A)=\\
\frac{\ell}{2}\left[4\sqrt{\frac{\gamma t}{\pi\ell^2}}
\left(e^{-\ell^2/(16\gamma
t)}-1\right)+\mathrm{Erf}\left(\frac{\ell}{4\sqrt{\gamma
t}}\right)\right]\\
\times\int\!\!\frac{dk}{2\pi}\left[
{\mathcal F}(\lambda_+)+{\mathcal{F}(\lambda_-)}-{\mathcal{F}(f(k)/2)}-{\mathcal{F}(f(k+\pi)/2)}\right]
\\+\frac{\ell}{2}\int\frac{dk}{2\pi}[{\mathcal{F}}(f(k)/2)+{\mathcal{F}(f(k+\pi)/2)}].  
\end{multline}
It is convenient to rewrite Eq.~\eqref{eq:trfga} as 
\begin{multline}
	\label{eq:trfga-1}
	\mathrm{Tr}{\mathcal F}(G_A)=\frac{1}{2}
	\int_{-\infty}^\infty \!\!\!\! d\xi P(\xi)\int\frac{dk}{2\pi}
	\Big\{ \\\max[\ell-2|\xi|,0]
[{\mathcal F}(\lambda_+)
+{\mathcal F}(\lambda_-)]\\+
\min(2|\xi|,\ell)\left[{\mathcal F}\left(f(k)/2\right)
+	{\mathcal F}\left(f\left(k+\pi\right)/2\right)\right]\Big\}. 
\end{multline}
Here we defined $P(\xi,t)$ as 
\begin{equation}
	\label{eq:px}
	P(\xi,t)=\frac{1}{2\sqrt{\pi\gamma t}}e^{-\xi^2/(4\gamma t)}. 
\end{equation}
Eq.~\eqref{eq:trfga-1} admits a simple hydrodynamic interpretation in terms of 
quasiparticles. As in the standard quasiparticle picture for entanglement spreading, the starting 
point is that entangled pairs of quasiparticles are produced after the quench. 
Unlike in the deterministic chain, quasiparticles spread diffusively, and their trajectories 
do not depend on $k$. In~\eqref{eq:trfga-1} $\xi$ is interpreted as 
the distance of the quasiparticles from the point at which they were initially 
emitted. Eq.~\eqref{eq:trfga-1} suggests that the two 
members of the entangled pairs perform symmetric (``mirrored'') random walks with respect to the emission 
point (see Fig.~\ref{fig:intro}). The fact that at any given time 
the hopping amplitudes do not depend on the position suggests that 
all the pairs undergo the same dynamics, for each realization of the noise. 
Thus, the term $\max(\ell-2|\xi|,0)$ in~\eqref{eq:trfga-1} counts the number of entangled quasiparticles  
that at time $t$ travelled a distance $|\xi|$ and are within $A$, whereas $\min(2|\xi|,\ell)$ is 
the number of shared entangled pairs, i.e., such that 
at time $t$ only one member of the pair is in $A$. Although 
the kinematic structure of~\eqref{eq:trfga-1} is the same 
as in the deterministic chain~\cite{calabrese2012quantum,alba2022hydrodynamics,alba2023logarithmic}, 
the dynamics is stochastic, and  $\xi$ is a random variable distributed with $P(x,t)$ (cf.~\eqref{eq:px}). 
Finally, the contributions $\mathcal{F}(\lambda_+)+\mathcal{F}(\lambda_-)$ and 
$\mathcal{F}(f(k)/2)+\mathcal{F}(f(k+\pi)/2)$ 
of the pairs to $\mathrm{Tr}\mathcal{F}(G_A)$ are the same as in the deterministic chain. 
One can check that by replacing $P(\xi)\to \delta(\xi-v_kt)$, with $v_k$ the group 
velocity of the quasiparticles, one recovers the result for the quench in the deterministic chain. 

From~\eqref{eq:trfga-1} one obtains the dynamics of the R\'enyi entropies and the von Neumann entropy 
by choosing $\mathcal{F}(x)$ as 
\begin{align}
	\label{eq:f1}
	& {\mathcal F}_{n}(x)=\frac{1}{1-n}\ln\left[x^n+(1-x)^n\right]\\
	\label{eq:f2}
	& {\mathcal F}_1(x)=-x\ln(x)-(1-x)\ln(1-x), 
\end{align}
respectively. 
%
%############################################
\begin{figure}[t]
	\begin{center}
\includegraphics[width=.9\linewidth]{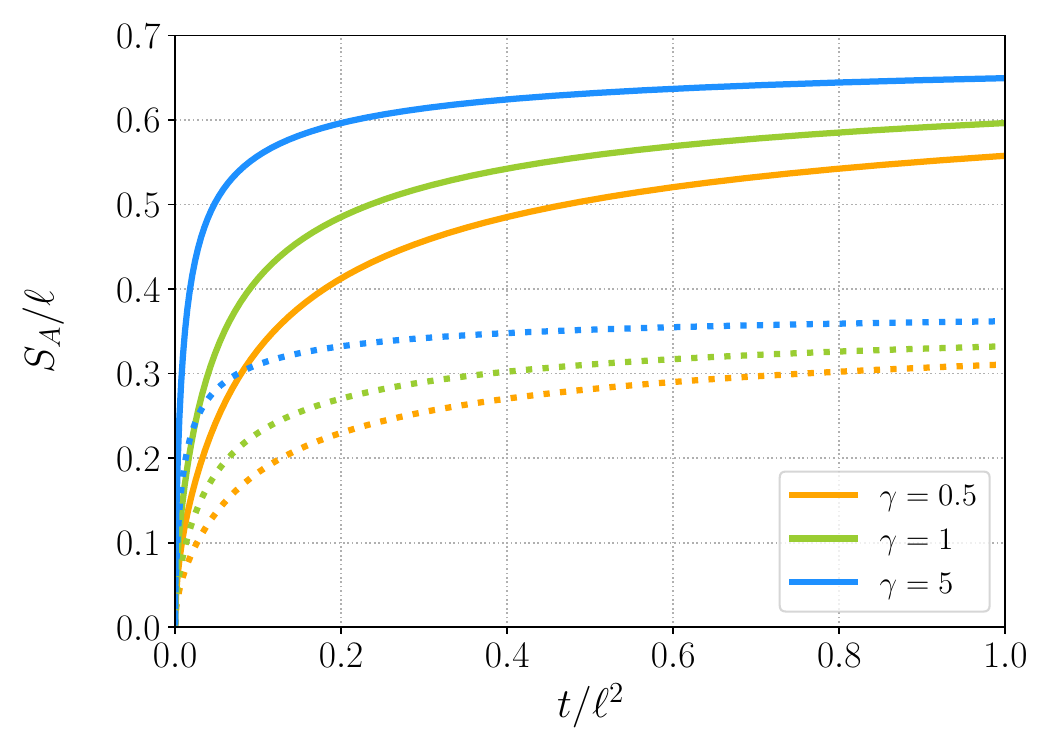}
\caption{ Entanglement dynamics in the $\nu$-$QSSEP$ with $\nu=1$ after 
	the quench from the fermionic N\'eel state (continuous line) 
	and the dimer state (dotted line). We show results for several 
	values of the noise rate $\gamma$, plotting $S_A/\ell$ versus 
	$t/\ell^2$, with $\ell$ the size of the subsystem. 
}
\label{fig:theo-ent}
\end{center}
\end{figure} 
%############################################
%
For~\eqref{eq:f1} and~\eqref{eq:f2} ${\mathcal F}(1)={\mathcal{F}(0)}=0$, 
which reflects that the full-system entropies are zero because the state is 
pure for each realization of the hopping amplitudes. 
Thus, for the entropies Eq.~\eqref{eq:neel-Mn} can be rewritten as 
\begin{equation}
	\label{eq:ent-quasi}
	S_n=\int_{-\infty}^\infty d \xi \min(2|\xi|,\ell) s_n P(\xi), 
\end{equation}
where 
\begin{equation}
	s_n=\int \frac{dk}{2\pi} {\mathcal F}_n\left(\frac{f(k)}{2}\right)
\end{equation}
Eq.~\eqref{eq:ent-quasi} provides a stochastic generalization of the 
quasiparticle picture for entanglement spreading. 

In Fig.~\ref{fig:theo-ent} we plot Eq.~\eqref{eq:ent-quasi} for the quench 
from the N\'eel state and the dimer state (cf.~\eqref{eq:neel} and~\eqref{eq:dimer}). 
The results are reported with the continuous lines and the dotted lines, respectively. 
For the N\'eel quench the R\'enyi entropy shared by the members of the entangled pairs is  
$s_n=\ln(2)$ for any $n$, whereas for the quench from the dimer state one has~\cite{alba2023logarithmic} 
\begin{equation}
	\label{eq:ent-dim}
	s_n=
	\frac{1}{1-n}\ln\Big[\Big(\frac{1+\cos(k)}{2}\Big)^n+
		\Big(\frac{1-\cos(k)}{2}\Big)^n
	\Big]. 
\end{equation}
The von Neumann entropy $s_1$ is obtained by taking the limit $n\to1$. 
In Fig.~\ref{fig:theo-ent} we plot the von Neumann entropy $S/\ell$ 
versus $t/\ell^2$, with $\ell$ the size of interval $A$ (see 
Fig.~\ref{fig:intro}). The growth of $S_A$ is diffusive for both quenches and 
for any $\gamma$. The saturation value for the quench from the dimer state 
is $S_A/\ell\to \ln(2)/2$, whereas for the N\'eel quench one has $\ln(2)$. 
Although the asymptotic value of the von Neumann entropy at 
long times does not depend on $\gamma$, Fig.~\ref{fig:theo-ent} 
shows that it is approached faster for larger $\gamma$.

%############################################
\section{Numerical results}
\label{sec:numerics} 

Here we discuss numerical results for the dynamics in 
the $\nu$-$QSSEP$, focussing on the case with $\nu=1$. 
In Section~\ref{sec:num-0} we detail the numerical strategy to 
simulate the full-time dynamics.  
We discuss the dynamics of the 
moments $M_n=\mathrm{Tr}(G_A^n)$ in Section~\ref{sec:num-1}. 
Then we provide numerical evidence supporting the validity 
of the stochastic quasiparticle picture~\eqref{eq:ent-quasi} 
for the von Neumann entropy after the 
quench from the N\'eel state and the dimer state in Section~\ref{sec:num-2}. 

%############################################
\subsection{Numerical method}
\label{sec:num-0} 

Here we employ the strategy of Ref.~\cite{cao2019entanglement}. 
We generate the hopping amplitudes $dW_t$ as $dW^r_t+idW^i_t$ with $dW_t^r$ and 
$dW_t^i$ real random numbers distributed with a Gaussian distribution with zero mean 
and variance $\gamma dt/2$. The generic initial state is represented as 
\begin{equation}
	\label{eq:in-U}
	|\psi_0\rangle=\prod_{k=1}^N\sum_{j=1}^L U_{jk}^{(0)}c^\dagger_j|0\rangle, 
\end{equation}
with  $U^{(0)}_{jk}$ a $L\times N$ matrix, and $N$ the number of fermions in the chain. 
We require that $[U^{(0)}]^\dagger U=\mathds{1}_{N}$, with $\mathds{1}_N$ the $N\times N$ identity matrix. 
One can repeatedly apply the infinitesimal evolution operator $e^{-idH_t}$, with $dH_t$ given in~\eqref{eq:ham}. 
The state~\eqref{eq:in-U} keeps the same form as in~\eqref{eq:in-U} during the dynamics.  
The evolved $U^t$ is obtained by first calculating $W^{t+dt}=e^{-idh_t}U^{t}$, 
with $dh_t$ given in~\eqref{eq:ham-mat}. To conclude the iteration step we perform a QR decomposition 
as $W^{t+dt}=QR$, and set $U^{t+dt}=Q$. Notice that the last step is not needed here because the 
dynamics, even for finite $dt$, preserves the condition $U^\dagger U=\mathds{1}_N$. The correlation matrix 
$C_{ij}$ is obtained as~\cite{cao2019entanglement} 
\begin{equation}
	\label{eq:Cij}
	C_{ij}=\langle \psi_t|c^\dagger_i c_j|\psi_t\rangle= U^t[U^t]^\dagger. 
\end{equation}
In our simulations we employ $dt=0.05$. We checked that the numerical results do not change significantly 
if we consider smaller $dt$. Finally, from the correlation function~\eqref{eq:Cij} we obtain 
entanglement-related quantities via the Peschel trick~\cite{peschel2009reduced}.

%############################################
\subsection{Moments $M_n$}
\label{sec:num-1} 

%
%############################################
\begin{figure}[t]
\centering
\includegraphics[width=.9\linewidth]{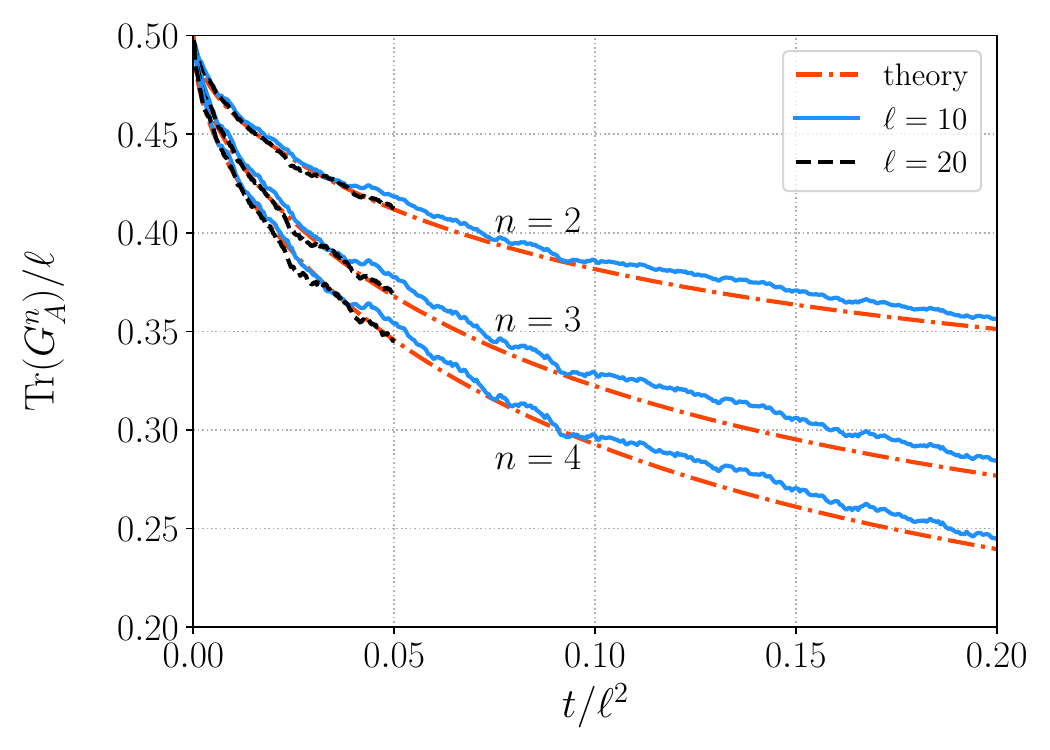}
\caption{ Moments $M_n=\mathrm{Tr}(G_A^n)$ after the 
	quench from the N\'eel state in the $\nu$- $QSSEP$ 
	with $\nu=1$ and $\gamma=0.5$. We plot $M_n/\ell$ 
	versus $t/\ell^2$, for $\ell=20,40,80$. 
	The continuous lines are exact numerical data averaged 
	over $\sim 500$ noise realizations. The dashed-dotted 
	lines are the theoretical predictions. 
}
\label{fig:trneel}
\end{figure} 
%############################################
%

In Fig.~\ref{fig:trneel} we plot $M_n$ with 
$n=2,3,4$ for the quench from the N\'eel state. The continuous and 
the dotted lines are exact numerical data for $\ell=10,20$, respectively. 
The data are for a chain with $L=200$ and $M_n$ are averaged over $500$ realizations 
of the hopping amplitudes $dW_t^j$. We checked that boundary effects due to the 
finite $L$ are negligible, at least for the times $t/\ell^2\approx 0.2$ that we consider. 
All the data are for fixed $\gamma=1/2$. In the Figure we plot $M_n/\ell$ versus $t/\ell^2$. At $t=0$ one has 
$M_n=1/2$ for any $n$. At long times $M_n$ decay to zero. The dash-dotted 
line is the prediction~\eqref{eq:neel-Mn}, which holds in the limit $\ell,t\to\infty$ with $t/\ell^2$ fixed. 
Even for $\ell=10,20$, finite-size effects are mild, and the agreement between the numerical 
data and the asymptotic result is perfect. 
%
%############################################
\begin{figure}[t]
\centering
\includegraphics[width=.9\linewidth]{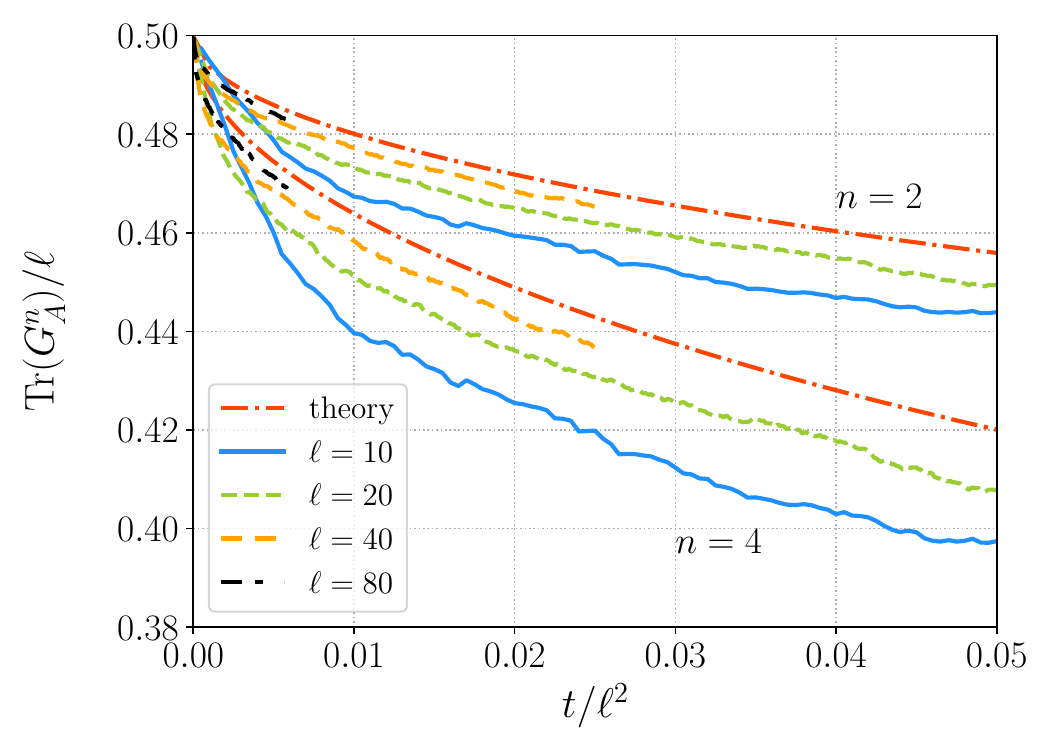}
\caption{ Same as in Fig.~\ref{fig:trneel} for the quench from 
	the dimer state. Finite-size effects are stronger than 
	in the N\'eel quench. 
}
\label{fig:trdimer}
\end{figure} 
%############################################
%

Let us now discuss the quench from the dimer state~\eqref{eq:dimer}. 
We show numerical data for $M_2$ and $M_4$ in Fig.~\ref{fig:trdimer}. 
The data are for fixed $\gamma=1/2$ and for a chain with $L=200$. 
In contrast with the quench from the N\'eel state (see Fig.~\ref{fig:trneel}) 
we observe sizeable finite-size corrections. The qualitative behavior is 
similar as for the N\'eel quench. At $t=0$  one has 
$M_n=1/2$, and $M_n$ decrease upon increasing $t$. To understand the 
corrections we plot data for several subsystem sizes up to $\ell=80$. 
Upon increasing $\ell$ the data approach the asymptotic result 
(dashed-dotted line in the figure). Already the result for $\ell=80$ 
is quite close to the asymptotic result. A fit of the data 
at fixed $t/\ell^2$ suggests that corrections are compatible with a 
$1/\ell$ decay, which is the same behavior as in the deterministic chain. 

%############################################
\subsection{von Neumann entropy}
\label{sec:num-2} 

\begin{figure}[t]
\centering
\includegraphics[width=.9\linewidth]{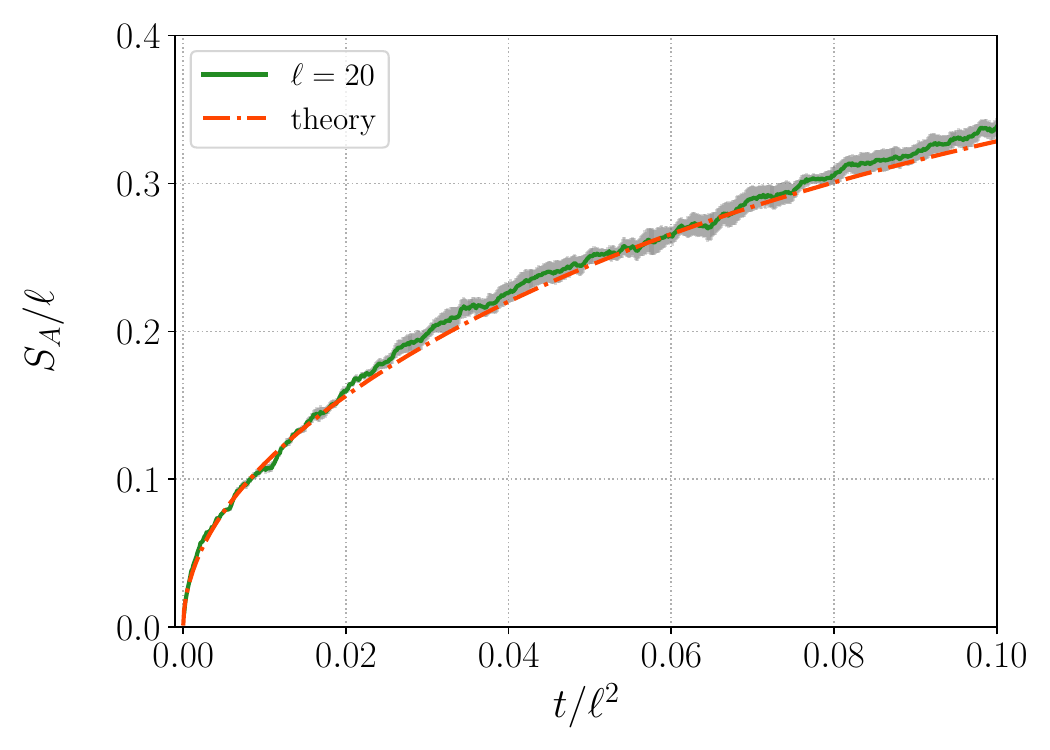}
\caption{ Dynamics of the von Neumann entropy of a subsystem of 
	length $\ell$ plotted as a function of $t/\ell^2$ in 
	the $\nu$-$QSSEP$ with $\nu=1$ and noise strength $\gamma=0.5$. 
	The dynamics is from the N\'eel state. The continuous and the dotted lines 
	are numerical results averaged over $1000$ realizations 
	of the dynamics. The dashed-dotted line is the analytic result 
	in the limit $\ell,t\to\infty$, with the ratio $t/\ell^2$. 
}
\label{fig:neel-ent}
\end{figure}

Let us now discuss the von Neumann entropy. In Fig.~\ref{fig:neel-ent} 
we plot $S_A/\ell$ versus $t/\ell^2$ for the quench from the N\'eel state. 
As in Fig.~\ref{fig:trneel} the data are for fixed $\gamma=1/2$. 
The continuous line is the exact numerical result for $S_A$ for 
$\ell=20$. The gray shading is the standard deviation of the mean 
computed by averaging the data over the noise realizations. 
The dashed-dotted line is~\eqref{eq:ent-quasi} with 
$s_n=\ln(2)$. As we observed for $M_n$ (see Fig.~\ref{fig:trneel}), 
the agreement with the theory is impressive, despite the ``small'' 
$\ell=20$. 
%
%############################################
%############################################
%

%
%############################################
\begin{figure}[t]
\centering
\includegraphics[width=.9\linewidth]{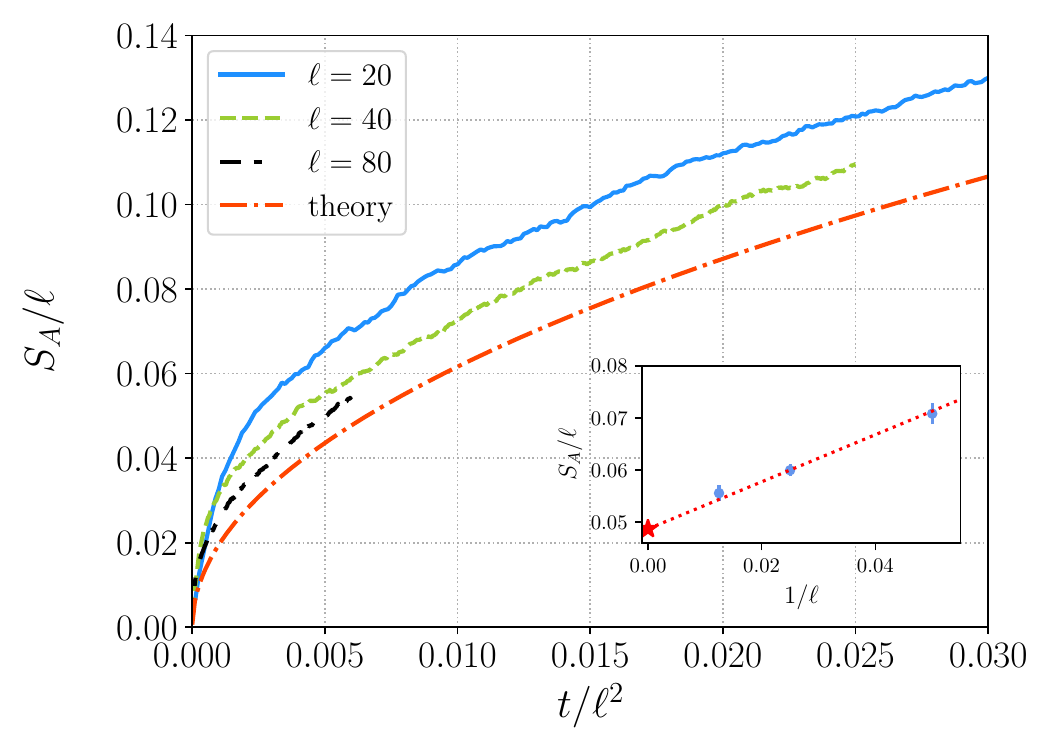}
\caption{ The same as in Fig.~\ref{fig:neel-ent} for the 
	quench from the dimer state. Finite-size corrections 
	are larger compared to the dynamics from the N\'eel state 
	(see Fig.~\ref{fig:neel-ent}). In the inset we show 
	the data for $S_A/\ell$ at fixed $t/\ell^2=0.625$ plotted 
	as a function of $1/\ell$. The star symbol is the analytic 
	result in the limit $\ell\to\infty$. The dotted line is  
	a fit to $a+b/\ell$, with $a$ fixed and $b$ a fitting parameter. 
}
\label{fig:dimer-ent}
\end{figure} 
%############################################
%

Fig.~\ref{fig:dimer-ent} shows results for $S_A$ for the quench from the 
dimer state~\eqref{eq:dimer}. Deviations between the exact numerical 
data and the theoretical prediction in the large $\ell,t$ limit are larger compared with 
the N\'eel quench. 
The finite-size corrections are investigated in the inset by plotting $S_A/\ell$ 
versus $1/\ell$ for fixed $t/\ell^2=0.00625$. The star symbol at $\ell\to\infty$ 
is the theoretical prediction. The dotted line is a fit to $a+b/\ell$ with 
$a$ fixed and $b$ a fitting parameter. 

%
%############################################
\begin{figure}[t]
\centering
\includegraphics[width=.9\linewidth]{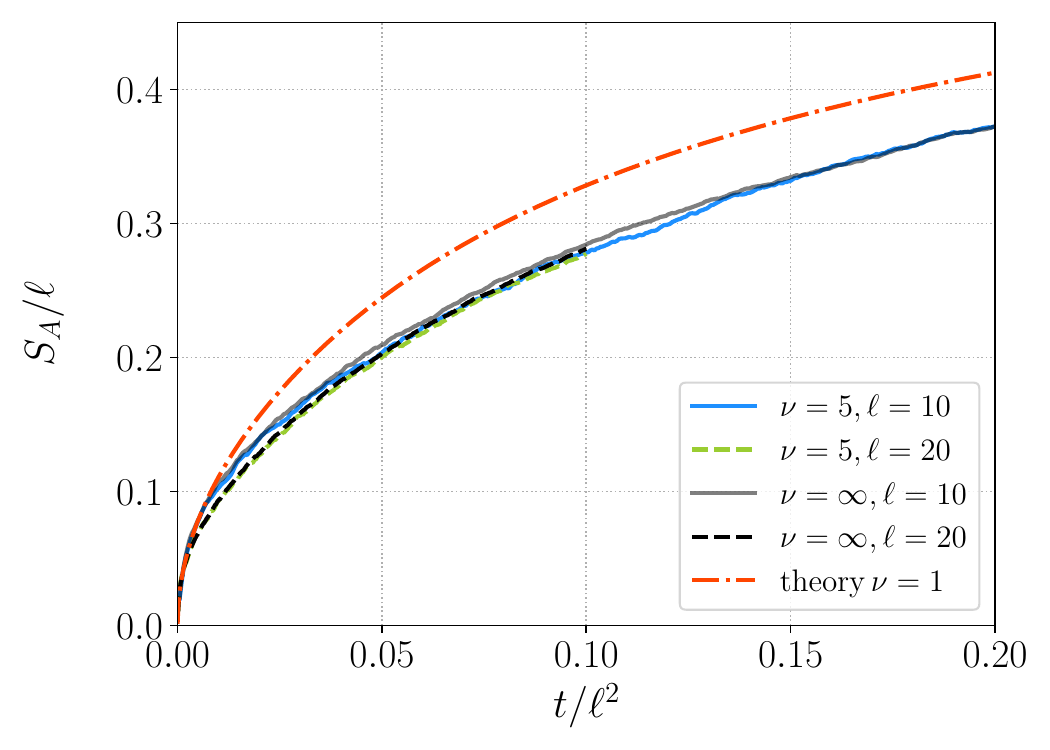}
\caption{ Entanglement dynamics after the quench from the N\'eel state in 
	the $\nu$-QSSEP model at noise rate $\gamma=0.5$. The continuous and 
	the dashed lines are exact numerical data for $\nu=5$ and $\ell=10$ and 
	$\ell=20$, respectively. The dashed-dotted line is the analytical result 
	for $\nu=1$ in the large $\ell,t$ limit. 
}
\label{fig:ent-nu}
\end{figure} 
%############################################
%
Finally, it is interesting to investigate the dependence on $\nu$ of the 
entanglement dynamics. In Fig.~\ref{fig:ent-nu} we consider the quench 
from the N\'eel state in the $\nu$-$QSSEP$ for fixed $\gamma=1/2$ and 
$\nu=5$ and $\nu\to\infty$. We show results for $\ell=10$ (continuous lines) 
and $\ell=20$ (dashed lines). The dashed-dotted line is the theoretical result~\eqref{eq:ent-quasi} 
for $\nu=1$. Interestingly, the data for $\nu=5$ and $\nu\to\infty$ for both 
$\ell=10$ and $\ell=20$ exhibit a satisfactory collapse, which means that 
$\nu=5$ is a good approximation for the dynamics in the $QSSEP$, i.e., for $\nu\to \infty$, 
at least for the quench from the N\'eel state. On the other hand, Eq.~\eqref{eq:ent-quasi} 
captures correctly the qualitative behavior, although it is not quantitatively accurate. 

%############################################
\section{Conclusions}
\label{sec:concl}

We investigated the out-of-equilibrium dynamics in a minimal modification of 
the so-called Quantum Symmetric Simple Exclusion Process ($QSSEP$), which is a tight-binding 
fermionic chain with space-time random complex hopping amplitudes. 
Here we considered the $\nu$-$QSSEP$ in which the hopping amplitudes retain their 
randomness in time but are defined in a unit cell of $\nu$ sites. 
We derived the dynamics of the generic moment of the fermionic correlation function in momentum 
space. We showed that the statistics of the moments in the steady-state  is invariant  
under the  Haar average with structured random matrices. The latter are 
obtained in terms of momentum-dependent $SU(\nu)$ random matrices. 
As a consequence, the moments of the fermionic correlators satisfy an intriguing set of relationships that 
we characterized in detail for $\nu=2$. 
For $\nu=1$ we characterized analytically the full-time dynamics of entanglement-related 
quantities, in the space-time scaling limit of long times and large subsystem 
sizes. In contrast with quantum quenches in the deterministic tight-binding chain, 
the entanglement spreading is diffusive for any $\nu$. 
For $\nu=1$ the entanglement dynamics is quantitatively described by 
a stochastic generalization of the quasiparticle picture. 
In this modified quasiparticle picture pairs of entangled qusiparticles are 
produced at each point in space. The two members of the pair undergo two symmetric 
random walks in opposite directions. All the entangled pairs produced in the system 
undergo the same dynamics. The distance from the emission point of each pair 
is random, and it is distributed with the heat kernel with diffusion constant $\gamma$, 
i.e., the strength of the hopping rate.  

Let us now discuss some possible directions for future work. First, here we analyzed in 
detail only the dynamics for $\nu=1$ and the steady-state for $\nu=2$. Clearly, it would be 
interesting to investigate the full-time dynamics of the entanglement entropy for 
$\nu>1$. For moderate values of $\nu$, for instance for $\nu=2$, this should be feasible, 
and it would allow to understand how to generalize the stochastic quasiparticle picture to 
generic $\nu$. The key ingredients are the eigenvalues $\varepsilon_{\{k_jq_j\}}$ governing the 
dynamics of the moments of the correlation functions. Although already for the second moment 
determining the exact eigenvalues requires diagonalizing~\eqref{eq:M6}, which is $6\times 6$, this 
should be possible in the limit in which all the momenta become equal, which is expected to 
govern the long-time dynamics. Furthermore, here we considered dynamics from low-entanglement 
initial states, with almost diagonal correlation functions in momentum space. It would be 
interesting to explore dynamics from initial states with more complex correlations. Moreover, 
it should be possible to generalize the stochastic quasiparticle picture to the dynamics of 
other entanglement-related quantities, such as the entanglement negativity~\cite{alba2019quantum}, or 
even the entanglement Hamiltonian~\cite{rottoli2025entanglement}. Finally, 
Eq.~\eqref{eq:ham} can be generalized to the interacting case. Although the interacting 
model is unlikely to be solvable, the dynamics could be studied by employing the 
time-dependent Density Matrix Renormalization Group~\cite{paeckel2019time} ($tDRMG$). It would be interesting 
to investigate whether interactions modify the diffusive nature of entanglement 
spreading, and whether they give rise to entanglement phase transitions.

\section*{Acknowledgements}

I would like to thank Bruno Bertini and Lorenzo Piroli for useful discussions. 
I would also like to thank Angelo Russotto, Filiberto Ares, and Pasquale Calabrese for 
discussions on a related topic. 
This study was carried out within the National Centre on HPC, Big Data and 
Quantum Computing - SPOKE 10 (Quantum Computing) and received funding 
from the European Union Next- GenerationEU - National Recovery and 
Resilience Plan (NRRP) – MISSION 4 COMPONENT 2, 
INVESTMENT N. 1.4 – CUP N. I53C22000690001. This work has been 
supported by the project “Artificially devised many-body 
quantum dynamics in low dimensions - ManyQLowD” funded by the 
MIUR Progetti di Ricerca di Rilevante Interesse Nazionale
(PRIN) Bando 2022 - grant 2022R35ZBF. 

\bibliography{bibliography}

%\appendix

%\bibliography{bibliography.bib} \nolinenumbers

\appendix

%############################################
\section{Detailed derivation of $M_n$ for small $n$} 
\label{app:1}

Here provide details on the derivation of the large $\ell,t$ 
behavior of the moments $M_n$ of $G_A$ (cf.~\eqref{eq:moments}) 
for $n=2,3,4$. Let us consider the case of the quench from the N\'eel (cf.~\eqref{eq:neel})  
state (see Eq.~\eqref{eq:neel-Mn} for the result for $M_n$). 
For $M_2$ from~\eqref{eq:gen-form} we obtain that  
\begin{multline} 
\label{eq:2-si}
\langle \mathrm{Tr}(
G_A^2)\rangle\simeq
\frac{1}{4}\left(\frac{\ell}{4}\right)^2\int\frac{dk_1}{2\pi}\int\frac{dk_2}{2\pi}
\int_{-1}^1d\xi_1\int_{-1}^1d\xi_2\\ e^{-8\gamma t(1-\cos(k_1-k_2))
-i\ell(\xi_1-\xi_2)(k_1-k_2)/2}w^2(k_1-k_2).  
\end{multline}
Here we used that the only nonzero entries of $G_{k_1q_1}G_{k_2q_2}$ are for 
$q_j=k_j$ and $q_j=k_j\pm\pi$. Moreover, we used that in the large 
$\ell,t$ limit stationarity with respect $\xi_j$ implies that 
$k_1$ and $k_2$ are ``paired'', meaning that 
both $q_1$ and $q_2$ are shifted as $q_1=k_1\pm\pi$ and $q_2=k_2\mp\pi$. 
Other configurations, for instance with $q_1=k_1$ and $k_2=q_2$ are subleading in the 
limit $\ell\to\infty$. Actually, for the quench from the N\'eel state, since $G_{kq}$ 
does not depend on the quasimomenta, these term vanish exactly. This reflects that 
finite $\ell$ corrections are smaller for the quench from the N\'eel state. 
Finally, in~\eqref{eq:2-si} we used that $\langle G_{k_1k_1+\pi}G_{k_2k_2-\pi}\rangle(t)=
1/4 e^{-8\gamma t(1+\cos(k_1-k_2))}$. The phase factor $e^{i\ell\xi_1(q_2-k_1)/2+i\ell\xi_2(q_1-k_2)/2}$ 
would give $e^{i\ell\xi_1(k_2\pm\pi-k_1)/2+
i\ell\xi_2(k_1\mp\pi-k_2)/2}$ after using that $k_1$ and $k_2$ have to be paired.  
By changing variable $k_1\to k_1\pm \pi$ one obtains~\eqref{eq:2-si}.  
Notice that in the exponent in~\eqref{eq:2-si} one has  
$-8\gamma t (1-\cos(k_1-k_2))$ instead of$-8\gamma t (1+\cos(k_1-k_2))$   
(cf.~\eqref{eq:vel-1}), due to the change of variables. 

Moreover, the $\simeq$ in~\eqref{eq:2-si} is 
to stress that we are considering only the terms 
that evolve with time, ignoring the non-evolving ones, which correspond to 
$k_1=q_1,k_2=q_2$, $q_1=k_2,q_2=k_1$, and $k_1=k_2,q_1=q_2$. 
By changing variables as $\zeta_0=\xi_1$ and $\zeta_1=\xi_1-\xi_2$, we obtain 
\begin{multline}
	\label{eq:2-si-1}
	\langle \mathrm{Tr}(\langle G_A^2)\rangle\simeq
	\frac{1}{4}\left(\frac{\ell}{4}\right)^2\int\frac{dk_1}{2\pi}\int\frac{dk_2}{2\pi}
	\int_{-2}^2d\zeta_1\\ e^{-8\gamma t(1-\cos(k_1-k_2))
	-i\ell\zeta_1(k_1-k_2)/2}w^2(k_1-k_2) \mu(\zeta_1).  \end{multline}
where we used that the integrand does not depend on $\zeta_0$. One should observe that 
a crucial difference with the dynamics under the deterministic tight-binding chain~\cite{alba2023logarithmic} 
is that the momenta $k_1,k_2$ are coupled in a nonlinear way by the $\cos(k_1-k_2)$ term. This gives 
rise to diffusive entanglement spreading. 
The integration over $\zeta_0$ gives  $\mu(\zeta_1)$, which is defined as 
\begin{equation} \label{eq:mu}
	\mu(\zeta_1)=\max(0,\min(1,1-\zeta_1)+\min(1,1+\zeta_1)).
\end{equation}
Crucially, in the large $\ell,t$ limit one can imagine of treating the integral 
in~\eqref{eq:2-si-1} via the stationary phase approximation~\cite{wong2001asymptotic}. 
The stationarity condition with respect to 
$\zeta_1$ implies that only the region $k_1\approx 
k_2$ contributes in the integral. This allows us to expand the $\cos(k_1-k_2)$, and
replace $w(k_1-k_2)\to2$.  We obtain 
\begin{multline} \langle\mathrm{Tr}(G_A^2)\rangle\simeq 
\left(\frac{\ell}{4}\right)^2\int\frac{dk_1}{2\pi}\int\frac{dk_2}{2\pi}
\int_{-2}^2 d\zeta_1\\ \mu(\zeta_1)e^{-4\gamma t(k_1-k_2)^2
-i\ell\zeta_1(k_1-k_2)/2}, \end{multline}
We can perform the integral over $k_1$, which gives 
\begin{multline} \label{eq:stepx} \langle\mathrm{Tr}(G_A^2)\rangle\simeq
	\left(\frac{\ell}{4}\right)^2\int\frac{dk_2}{2\pi} \int_{-2}^2 d\zeta_1
	\mu(\zeta_1)\\
	\frac{i}{8\sqrt{\pi\gamma t}}e^{-\ell^2\zeta_1^2/(64\gamma
	t)}\left[ \mathrm{Erfi}\left(\frac{\ell\zeta_1+16i\gamma t
	(k_2-\pi)}{8\sqrt{\gamma t}}\right)\right.\\-\left.
\mathrm{Erfi}\left(\frac{\ell\zeta_1+16i\gamma t (k_2+\pi)}{8\sqrt{\gamma
t}}\right) \right], \end{multline}
where $\mathrm{Erfi}(x)=-i \mathrm{Erf}(i x)$, with  $\mathrm{Erf}(x)$ 
the error function. 
It is now clear that the dynamics is diffusive since the combination $\ell^2/t$ appears 
in~\eqref{eq:stepx}. 
We can simplify Eq.~\eqref{eq:stepx} by taking the limit $\ell,t\to\infty$ with
$\ell/\sqrt{t}$ fixed to extract the leading behavior of~\eqref{eq:stepx}.  
In this limit the dependence on $k_2$ disappears, and  we obtain 
\begin{multline} \langle\mathrm{Tr}(G_A^2)\rangle\simeq 
\left(\frac{\ell}{4}\right)^2\!\!\!\int\!\!\frac{dk_2}{2\pi} 
\int\!\! d\zeta_1
\mu(\zeta_1) \frac{e^{-\ell^2\zeta_1^2/(64\gamma t)}}{4\sqrt{\pi\gamma t}}.
\end{multline}
Finally, the integration over $k_2$ is trivial, and the integral over $\zeta_1$ gives  
\begin{multline} \label{eq:tr-2-res} \langle\mathrm{Tr}(
G_A^2)\rangle\simeq\\
\ell\left[\sqrt{\frac{\gamma t}{\pi\ell^2}}
\left(e^{-\ell^2/(16\gamma
t)}-1\right)+\frac{1}{4}\mathrm{Erf}\left(\frac{\ell}{4\sqrt{\gamma
t}}\right)\right] \end{multline}
Let us now discuss the steady-state value of 
$\langle\mathrm{Tr}(G_A^n)\rangle$, which we have to add to~\eqref{eq:tr-2-res} to 
recover the full result. We have 
\begin{multline}
	\langle\mathrm{Tr}(G_A^2)\rangle =
	\frac{1}{L^2}
	\sum_{k_j,q_j}\sum_{n,m=1}^\ell \\
	e^{-i(k_1-q_2)n+i(q_1-k_2)m} G_{k_1q_1}(t)
	G_{k_2q_2}(t). 
\end{multline}
Now, the only terms attaining a nonzero value for $t\to\infty$ correspond to 
$k_1=q_1=k_2=q_2$, $k_1=q_1\ne k_2=q_2$, and $k_1=q_2\ne k_2=q_1$. Crucially, 
due to the fact that $G_{kq}$ is almost diagonal, reflecting the translation 
invariance of the initial states, one can check that only the terms with 
$k_1=q_1\ne k_2=q_2$ contribute at the leading order in the limit $L\to\infty$ 
and $\ell\to\infty$. For instance, the term with $k_1=q_2$ and $k_2=q_1$ would give 
\begin{equation}
	\label{eq:cc}
	\left(\ell/L\right)^2\sum_{k,q}G_{kq}G_{qk}\sim \frac{\ell^2}{L}, 
\end{equation}
Now, Eq.~\eqref{eq:cc} vanishes if one takes the thermodynamic limit $L\to\infty$ first 
and then the limit $\ell\to\infty$. 
One can check that for translation invariant states, such as the N\'eel state and the 
dimer state, this generalizes to arbitrary $n$. Precisely, the steady-state value of 
$M_n$ is determined by the ``diagonal'' terms $G_{k_1k_1}G_{k_2k_2}\cdots G_{k_nk_n}$. 
This means that 
\begin{equation}
	\label{eq:steady}
	M_n\xrightarrow{L,t,\ell\to\infty}
	\ell\int_{-\pi}^\pi\frac{dk}{2\pi}G_{kk}^n, 
\end{equation}
where the limit $L,t,\ell\to\infty$ is taken in the specified order. 
Eq.~\eqref{eq:steady} implies that the steady-state value of $M_n$ 
is determined by the same Generalized Gibbs Ensemble ($GGE$) 
that describes the steady state after the quench with the deterministic 
tight-binding Hamiltonian. 

For the quench from the N\'eel state one has $G_{kk}=1/2$. 
From~\eqref{eq:tr-2-res} we obtain 
\begin{multline}
M_2=\ell\Big[\sqrt{\frac{\gamma t}{\pi\ell^2}}
\left(e^{-\ell^2/(16\gamma
t)}-1\right)\\+\frac{1}{4}\mathrm{Erf}\left(\frac{\ell}{4\sqrt{\gamma
t}}\right)\Big] + \frac{\ell}{4}
\end{multline}
where $\ell/4$ is the steady-state contribution.

Let us now consider the case of $M_3$. We now have to consider the average 
$\langle G_{k_1q_1}G_{k_2q_2}G_{k_3q_3}\rangle$. Again, let us consider only 
the time-dependent terms in the dynamics, i.e., neglecting the steady-state 
value of $M_3$. Now, from~\eqref{eq:Gk-evol} and~\eqref{eq:evol-square} we have 
\begin{equation}
	\frac{d\langle G_{k_1q_1}G_{k_2q_2}G_{k_3q_3}\rangle}{dt}=\varepsilon_{\{k_jq_j\}}
	\langle G_{k_1q_1}G_{k_2q_2}G_{k_3q_3}\rangle. 
\end{equation}
Now if there is only one $\pi$ shift  $q_j=k_j\pm \pi$ for some $j\in[1,3]$ 
one has $\varepsilon=-4$. The configuration with two shifts as $q_j=k_j\pm \pi$ and $q_{j'}=k_{j'}\pm
\pi$ corresponds to energy $\varepsilon=-8(1+\cos(k_j-k_{j'}))$. If all the $q_j$ are shifted one has
$\varepsilon=-2(6+4\cos(k_1-k_2)+4\cos(k_1-k_3)+4\cos(k_2-k_3))$. Again, 
we observe that in the large $\ell$ limit, within the stationary phase approximation, 
only the terms with $q_j=k_j\pm \pi$ and $q_{j'}=k_{j'}\mp\pi$, i.e., with $k_j,k_{j'}$ paired,  
contribute. We have 
\begin{multline} \label{eq:tr-3-int} 
	M_3\simeq\left(\frac{\ell}{4}\right)^3 \int\!\!\frac{dk_1}{2\pi}\int\!\!
	\frac{dk_2}{2\pi}\int\!\!\frac{dk_3}{2\pi}\int\!\! d\xi_1\int\!\! d\xi_2\int\!\! 
	d\xi_3
	\\2^{-3}e^{i\ell(\xi_1-\xi_2)k_2/2+i\ell(\xi_2-\xi_3)k_3/2+i\ell(\xi_3-\xi_1)k_1/2}
	w(k_2-k_1)\\
	\times w(k_3-k_2)w(k_1-k_3) \sum_{i<j}e^{-8\gamma
t(1-\cos(k_i-k_j))}
\end{multline}
where the factor $2^{-3}$ originates from 
$G_{kq}=1/2\delta_{kq}\,\mathrm{mod}\,\pi$ for the 
N\'eel state. The terms in the sum in~\eqref{eq:tr-3-int} correspond to  
the three different pairings  of the momenta $q_j=k_j\pm \pi$, $q_{j'}=k_{j'}\mp\pi$. 
Similar to the case of $M_2$, we redefined the quasimomenta $k_j$ to remove the 
$\pi$ shifts from the exponent in the second row in~\eqref{eq:tr-3-int}. 
This also leads to the term  $1-\cos(k_i-k_j)$ in the exponent 
in the last row. 
Again, in the large $t,\ell$ limit only the region $k_i\approx k_j$ contributes 
in the integral, and we can expand the cosine functions, to obtain 
\begin{multline} 
	\label{eq:three}
	M_3\simeq\left(\frac{\ell}{4}\right)^3 \int\frac{dk_1}{2\pi}\int
	\frac{dk_2}{2\pi}\int\frac{dk_3}{2\pi}\int d\zeta_1\int d\zeta_2
	\mu(\{\zeta_j\})\\\times e^{i\ell\zeta_1 (k_1-k_3)/2+i\ell\zeta_2 (k_2-k_3)/2}
	\sum_{i<j=1}^3 e^{-4\gamma t(k_i-k_j)^2}, 
\end{multline}
where we also changed variables from the $\xi_j$ to the $\zeta_j$ (cf.~\eqref{eq:zeta-def}), 
we performed the integral with respect to $\zeta_0$, and replaced $w(k_i-k_j)\to 2$. 
Finally, one can check that the three terms in~\eqref{eq:three} give the same result, and we 
obtain that 
\begin{multline} M_3= 3\ell\Big[\frac{1}{2}\sqrt{\frac{\gamma t}{\pi\ell^2}}
\Big(e^{-\ell^2/(16\gamma
t)}-1\Big)\\
+\frac{1}{8}\mathrm{Erf}\Big(\frac{\ell}{4\sqrt{\gamma
t}}\Big)\Big]+\frac{\ell}{8},  \end{multline}
where the last term is the steady-state contribution at $t\to\infty$. 

Let us now consider $M_n$  with $n=4$. The dynamics of 
$\langle G_{k_1q_1}G_{k_2q_2}G_{k_3q_3}G_{k_4q_4}\rangle$ is governed by 
\begin{multline}
	d \langle G_{k_1q_1}G_{k_2q_2}G_{k_3q_3}G_{k_4q_4}\rangle =\\
	\varepsilon_{\{k_jq_j\}} \langle G_{k_1q_1}G_{k_2q_2}G_{k_3q_3}G_{k_4q_4}\rangle dt.  
\end{multline}
The energies $\varepsilon_{\{k_j,q_j\}}$ vanish if $k_j=q_j$
for all values of $j$.  If only one of the $q_j$ is shifted as $q_j=k_j\pm\pi$, then one has
$\varepsilon_{\{k_jq_j\}}=-4$. If one has $q_j=k_j\pm \pi$ and
${q_{j'}}=k_{j'}\pm\pi$ for some $j\ne j'$, then  
$\varepsilon_{\{k_jq_j\}}=-8(1+\cos(k_j-k_{j'}))$.  Finally, if  $q_j=k_j\pm \pi$ 
for all the values of $j$ one has $\varepsilon=-8(2+\sum_{j<j'}\cos(k_j-k_{j'}))$. 

Again, by employing the stationary phase approximation, one can verify that 
only the terms with an even number of paired momenta give a nonzero contribution in the 
limit $t,\ell\to\infty$. 
For $n=4$, one can have two  or four paired momenta. The case with two pairings 
corresponds to $q_j=k_j+\pi$ and $q_{j'}=k_{j'}-\pi$ for some $j\ne j'$. 
The terms with two pairings are treated as for $M_2$, and give all the same contribution. 
Let us consdier the case with four pairings focusing 
on the case with $q_{1,2}=k_{1,2}+\pi$, and  $q_{3,4}=k_{3,4}-\pi$. 
It is straightforward to check that  the result depends only on the number of pairings. 
Let us denote the contribution to $M_4$ originating from the case with four 
pairings as $M_4^{(\mathrm{4p})}$. $M_4^{(\mathrm{4p})}$ is given as  
\begin{multline} \label{eq:tr-4-last-2} 
	M_4^{(\mathrm{4p})}=\left(\frac{\ell}{4}
	\right)^4\prod_{j=1}^4\int\frac{dk_j}{2\pi}
	\int d\xi_j\Big\{
	e^{i\ell(\xi_1-\xi_4)k_4/2}\\
	e^{+i\ell(\xi_4-\xi_3)(k_3-\pi)/2+i\ell(\xi_3-\xi_2)k_2/2+i\ell(\xi_2-\xi_1)(k_1+\pi)/2}
\\G_{k_1,k_1+\pi}G_{k_2,k_2+\pi}G_{k_3,k_3-\pi}G_{k_4,k_4-\pi}\\
	w(k_4-\pi-k_1)w(k_1+\pi-k_2)w(k_2+\pi-k_3)w(k_3-\pi-k_4)\\ 
	e^{-8\gamma t[2+\cos(k_1-k_2)+\cos(k_1-k_3)+\cos(k_1-k_4)+\cos(k_2-k_3)]}\\
e^{-8\gamma t[\cos(k_2-k_4)+\cos(k_3-k_4)]}\Big\}
\end{multline}
Now, we can change variables by redefining  $k_1$ and $k_3$, 
to remove all the $\pi$ shifts. 
Let us also change variables to the $\zeta_j$ (cf.~\eqref{eq:zeta-def}). 
After integrating over $\zeta_0$, we obtain 
\begin{multline} \label{eq:tr-4-last-3} 
	M_4^{(\mathrm{4p})}= \left(\frac{\ell}{4}
	\right)^4\prod_{j=1}^4\int\frac{dk_j}{2\pi} \\
	\int d\zeta_1\int d\zeta_2\int d\zeta_3 \mu(\zeta_1,\zeta_2,\zeta_3)\Big\{
		2^{-4}
	e^{i\ell(k_3-k_4)\zeta_3/2}\\
	e^{+i\ell(k_2-k_4)\zeta_2/2+i\ell(k_1-k_4)\zeta_1/2}\\
	w(k_4-k_1)w(k_1-k_2)w(k_2-k_3)w(k_3-k_4)\\ e^{-8\gamma
t(2-\cos(k_1-k_2)+\cos(k_1-k_3)-\cos(k_1-k_4)-\cos(k_2-k_3)}\\
e^{-8\gamma t[\cos(k_2-k_4)-\cos(k_3-k_4)]}\Big\},
\end{multline}
Again, in the large $t,\ell$ limit we argue that only the 
region with all the $k_j$ equal  contributes to the 
integral, which allows us to expand the cosine functions 
in~\eqref{eq:tr-4-last-3}, and replace $w(k_i-k_j)\to 2$. 
The integration over $k_j$ is trivial, and it gives 
\begin{multline} \label{eq:tr-4-last-5} 
	M_4^{(\mathrm{4p})}= \left(\frac{\ell}{4} \right)^4 \int_{-2}^2
	d\zeta_1 d\zeta_2 d\zeta_3 \mu(\zeta_1,\zeta_2,\zeta_3)
	\frac{2^3}{\ell^3}\delta(\zeta_1+\zeta_2)\\
	\delta(\zeta_2+\zeta_3)\delta(\zeta_1-\zeta_3)
\frac{e^{-\ell^2\zeta_3^2/(64\gamma t)}}{8\sqrt{\pi\gamma t}} \end{multline}
In conclusion, after performing the integrals in $\zeta_j$, we obtain  
\begin{equation} \label{eq:tr-4-last-6} 
	M_4^{(\mathrm{4p})}=
	\frac{\ell}{2}\left[\frac{1}{2}\sqrt{\frac{\gamma t}{\pi\ell^2}}
	\left(e^{-\ell^2/(16\gamma
t)}-1\right)+\frac{1}{8}\mathrm{Erf}\left(\frac{\ell}{4\sqrt{\gamma
t}}\right)\right].  \end{equation}
Now, it is clear that the time-dependent term is of the same 
form as for $M_2$ and $M_3$, apart from the prefactors. 

\end{document}